%%%%%%%%%%%%%%%%%%%%%%%%%%
%% Toda hierarchy in super-Yang-Mills%
%% 17.12.2006 Hep-th 2nd Version %%%%%%%%%
%%% misprints corrected
%%%%%%%%%%%%%%%%%%%%%%%%%%

%%%%%%%%%%%%%%%%%%%%%%%%%%

%%%%%%%Russian fonts%%%%%%%5

\chardef\tempcat=\the\catcode`\@ \catcode`\@=11
\def\cyracc{\def\u##1{\if \i##1\accent"24 i%
    \else \accent"24 ##1\fi }}
\newfam\cyrfam
\font\tencyr=wncyr10
\newfam\cyrfam

\def\cyr{\fam\cyrfam\tencyr\cyracc}

% Pictures

\input epsf

\def\figin{\epsfcheck\figin}\def\figins{\epsfcheck\figins}
\def\epsfcheck{\ifx\epsfbox\UnDeFiNeD
\message{(NO epsf.tex, FIGURES WILL BE IGNORED)}
\gdef\figin##1{\vskip2in}\gdef\figins##1{\hskip.5in}% blank space instead
\else\message{(FIGURES WILL BE INCLUDED)}%
\gdef\figin##1{##1}\gdef\figins##1{##1}\fi}
\def\DefWarn#1{}
\def\figinsert{\goodbreak\topinsert}
\def\ifig#1#2#3#4{\DefWarn#1\xdef#1{fig.~\the\figno}
\writedef{#1\leftbracket fig.\noexpand~\the\figno}%
\figinsert\figin{\centerline{\epsfxsize=#3mm \epsfbox{#2}}}
\bigskip\medskip\centerline{\vbox{\baselineskip12pt
\advance\hsize by -1truein\noindent\footnotefont{\sl Fig.~\the\figno:}\sl\ #4}}
\bigskip\endinsert\noindent\global\advance\figno by1}

\def\Figx#1#2#3{
\bigskip
\vbox{\centerline{\epsfxsize=#1 cm \epsfbox{#2.eps}}
\centerline{{\bf Fig.\the\figno} #3}}\bigskip\global\advance\figno by1}

\def\Figy#1#2#3{
\bigskip
\vbox{\centerline{\epsfysize=#1 cm \epsfbox{NOpic#2.eps}}
\centerline{{\bf Fig.\the\figno} #3}}\bigskip\global\advance\figno by1}
\newcount\figno
 \figno=1
 \def\fig#1#2#3{
 \par\begingroup\parindent=0pt\leftskip=1cm\rightskip=1cm\parindent=0pt
 \baselineskip=11pt
 \global\advance\figno by 1
 \midinsert
 \epsfxsize=#3
 \centerline{\epsfbox{#2}}
 \vskip 12pt
 {\bf Fig.\ \the\figno: } #1\par
 \endinsert\endgroup\par
 }
 \def\figlabel#1{\xdef#1{\the\figno}}
 \def\encadremath#1{\vbox{\hrule\hbox{\vrule\kern8pt\vbox{\kern8pt
 \hbox{$\displaystyle #1$}\kern8pt}
 \kern8pt\vrule}\hrule}}

% Something to deal with sub-sub-sections

\def\unlockat{\catcode`\@=11}

\unlockat

% Sets global section and proposition numbers to zero
\global\newcount\secno \global\secno=0
\global\newcount\prono \global\prono=0
%%%%%%%%% Big sections
\def\newsec#1{\vfill\eject\global\advance\secno by1\message{(\the\secno. #1)}
\global\subsecno=0\global\subsubsecno=0
\global\deno=0\global\teno=0
\eqnres@t\noindent
{\titlefont\the\secno. #1}
\writetoca{{\bf\secsym} {\rm #1}}\par\nobreak\medskip\nobreak}
\global\newcount\subsecno \global\subsecno=0
%%%%%%%%%%%%%%%%Subsections %%%%%%%%%%%%%%%%%%%%%%%%%%%%%%%%%%%%%%%
\def\subsec#1{\global\advance\subsecno
by1\message{(\secsym\the\subsecno. #1)}
\ifnum\lastpenalty>9000\else\bigbreak\fi
\global\subsubsecno=0
\global\deno=0
\global\teno=0
%\eqnres@t
\noindent{\bf\secsym\the\subsecno. #1}
\writetoca{\bf \string\quad {\secsym\the\subsecno.} {\it  #1}}
\par\nobreak\medskip\nobreak}
\global\newcount\subsubsecno \global\subsubsecno=0
%%%%%%%%%%%%%%%%%%%%Subsubsections %%%%%%%%%%%%%%%%%%%%%%%%%%%%%
\def\subsubsec#1{\global\advance\subsubsecno by1
\message{(\secsym\the\subsecno.\the\subsubsecno. #1)}
\ifnum\lastpenalty>9000\else\bigbreak\fi
\noindent\quad{\bf \secsym\the\subsecno.\the\subsubsecno.}{\ \sl \ #1}
\writetoca{\string\qquad\bf { \secsym\the\subsecno.\the\subsubsecno.}{\sl  \ #1}}
\par\nobreak\medskip\nobreak}
%%% Definition

\global\newcount\deno \global\deno=0
\def\de#1{\global\advance\deno by1
\message{(\bf Definition\quad\secsym\the\subsecno.\the\deno #1)}
\ifnum\lastpenalty>9000\else\bigbreak\fi
\noindent{\bf Definition\quad\secsym\the\subsecno.\the\deno}{#1}
\writetoca{\string\qquad{\secsym\the\subsecno.\the\deno}{#1}}}
%%% Proposition

\global\newcount\prono \global\prono=0
\def\pro#1{\global\advance\prono by1
\message{(\bf Proposition\quad\secsym\the\subsecno.\the\prono %#1
)}
\ifnum\lastpenalty>9000\else\bigbreak\fi
\noindent{\bf Proposition\quad%\secsym\the\subsecno.
\the\prono\quad}{\ninepoint #1}
%\writetoca{\string\qquad{\secsym\the\subsecno.\the\prono}{#1}
}
%%% Theorem

\global\newcount\teno \global\prono=0
\def\te#1{\global\advance\teno by1
\message{(\bf Theorem\quad\secsym\the\subsecno.\the\teno #1)}
\ifnum\lastpenalty>9000\else\bigbreak\fi
\noindent{\bf Theorem\quad\secsym\the\subsecno.\the\teno}{#1}
\writetoca{\string\qquad{\secsym\the\subsecno.\the\teno}{#1}}}
%%%%%%%%%%%%
\def\subsubseclab#1{\DefWarn#1\xdef #1{\noexpand\hyperref{}{subsubsection}%
{\secsym\the\subsecno.\the\subsubsecno}%
{\secsym\the\subsecno.\the\subsubsecno}}%
\writedef{#1\leftbracket#1}\wrlabeL{#1=#1}}

\def\unredoffs{} \def\redoffs{\voffset=-.40truein\hoffset=-.40truein}
\def\speclscape{}

\newbox\leftpage \newdimen\fullhsize \newdimen\hstitle \newdimen\hsbody
\tolerance=1000\hfuzz=2pt

\catcode`\@=11
\def\bigans{b }
\def\answ{b }

\ifx\answ\bigans\message{(This will come out unreduced.}
\magnification=1200\unredoffs\baselineskip=16pt plus 2pt minus 1pt
\hsbody=\hsize \hstitle=\hsize

\else\message{(This will be reduced.} \let\l@r=L
\magnification=1200\baselineskip=16pt plus 2pt minus 1pt \vsize=7truein
\redoffs \hstitle=8truein\hsbody=4.75truein\fullhsize=10truein\hsize=\hsbody
\output={\ifnum\pageno=0

   \shipout\vbox{{\hsize\fullhsize\makeheadline}
     \hbox to \fullhsize{\hfill\pagebody\hfill}}\advancepageno
   \else
   \almostshipout{\leftline{\vbox{\pagebody\makefootline}}}\advancepageno
   \fi}
\def\almostshipout#1{\if L\l@r \count1=1 \message{[\the\count0.\the\count1]}
       \global\setbox\leftpage=#1 \global\let\l@r=R
  \else \count1=2
   \shipout\vbox{\speclscape{\hsize\fullhsize\makeheadline}
       \hbox to\fullhsize{\box\leftpage\hfil#1}}  \global\let\l@r=L\fi}
\fi

\newcount\yearltd\yearltd=\year\advance\yearltd by -2000

\def\Title#1#2{%\nopagenumbers
\abstractfont\hsize=\hstitle\rightline{#1}%
\vskip 5pt\centerline{\titlefont #2}\abstractfont\vskip .5in\pageno=0}
%

%&%

\def\draftmode{\message{ DRAFTMODE }\def\draftdate{{\rm preliminary draft:
\number\month/\number\day/\number\yearltd\ \ \hourmin}}%

\writelabels\baselineskip=20pt plus 2pt minus 2pt
  {\count255=\time\divide\count255 by 60 \xdef\hourmin{\number\count255}
   \multiply\count255 by-60\advance\count255 by\time
   \xdef\hourmin{\hourmin:\ifnum\count255<10 0\fi\the\count255}}}

\def\nolabels{\def\wrlabeL##1{}\def\eqlabeL##1{}\def\reflabeL##1{}}
\def\writelabels{\def\wrlabeL##1{\leavevmode\vadjust{\rlap{\smash%
{\line{{\escapechar=` \hfill\rlap{\sevenrm\hskip.03in\string##1}}}}}}}%
\def\eqlabeL##1{{\escapechar-1\rlap{\sevenrm\hskip.05in\string##1}}}%
\def\reflabeL##1{\noexpand\llap{\noexpand\sevenrm\string\string\string##1}}}
\nolabels
%

%\headline{\hfil\sl\dte}

\global\newcount\secno \global\secno=0
\global\newcount\meqno
\global\meqno=1
\def\eqnres@t{\xdef\secsym{\the\secno.}\global\meqno=1
\bigbreak\bigskip}
\def\sequentialequations{\def\eqnres@t{\bigbreak}}
\def\appendix#1#2{\vfill\eject\global\meqno=1\global\subsecno=0\xdef\secsym{\hbox{#1.}}
\bigbreak\bigskip\noindent{\bf Appendix #1. #2}\message{(#1. #2)}
\writetoca{Appendix {#1.} {#2}}\par\nobreak\medskip\nobreak}

\def\eqnn#1{\xdef #1{(\secsym\the\meqno)}\writedef{#1\leftbracket#1}%
\global\advance\meqno by1\wrlabeL#1}
\def\eqna#1{\xdef #1##1{\hbox{$(\secsym\the\meqno##1)$}}
\writedef{#1\numbersign1\leftbracket#1{\numbersign1}}%
\global\advance\meqno by1\wrlabeL{#1$\{\}$}}
\def\eqn#1#2{\xdef #1{(\secsym\the\meqno)}\writedef{#1\leftbracket#1}%
\global\advance\meqno by1$$#2\eqno#1\eqlabeL#1$$}

\newskip\footskip\footskip14pt plus 1pt minus 1pt

\def\footnotefont{\ninepoint}\def\f@t#1{\footnotefont #1\@foot}
\def\f@@t{\baselineskip\footskip\bgroup\footnotefont\aftergroup\@foot\let\next}
\setbox\strutbox=\hbox{\vrule height9.5pt depth4.5pt width0pt}
\global\newcount\ftno \global\ftno=0
\def\foot{\global\advance\ftno by1\footnote{$^{\the\ftno}$}}

\newwrite\ftfile
\def\footend{\def\foot{\global\advance\ftno by1\chardef\wfile=\ftfile
$^{\the\ftno}$\ifnum\ftno=1\immediate\openout\ftfile=foots.tmp\fi%
\immediate\write\ftfile{\noexpand\smallskip%
\noexpand\item{f\the\ftno:\ }\pctsign}\findarg}%
\def\footatend{\vfill\eject\immediate\closeout\ftfile{\parindent=20pt
\centerline{\bf Footnotes}\nobreak\bigskip\input foots.tmp }}}
\def\footatend{}

\global\newcount\refno \global\refno=1
\newwrite\rfile
\def\ref{[\the\refno]\nref}
\def\nref#1{\xdef#1{[\the\refno]}\writedef{#1\leftbracket#1}%
\ifnum\refno=1\immediate\openout\rfile=refs.tmp\fi \global\advance\refno
by1\chardef\wfile=\rfile\immediate \write\rfile{\noexpand\item{#1\
}\reflabeL{#1\hskip.31in}\pctsign}\findarg}

\def\findarg#1#{\begingroup\obeylines\newlinechar=`\^^M\pass@rg}
{\obeylines\gdef\pass@rg#1{\writ@line\relax #1^^M\hbox{}^^M}%
\gdef\writ@line#1^^M{\expandafter\toks0\expandafter{\striprel@x #1}%
\edef\next{\the\toks0}\ifx\next\em@rk\let\next=\endgroup\else\ifx\next\empty%
\else\immediate\write\wfile{\the\toks0}\fi\let\next=\writ@line\fi\next\relax}}
\def\striprel@x#1{} \def\em@rk{\hbox{}}
\def\lref{\begingroup\obeylines\lr@f}
\def\lr@f#1#2{\gdef#1{\ref#1{#2}}\endgroup\unskip}
\def\semi{;\hfil\break}
\def\addref#1{\immediate\write\rfile{\noexpand\item{}#1}}

\def\startrefs#1{\immediate\openout\rfile=refs.tmp\refno=#1}
\def\xref{\expandafter\xr@f}\def\xr@f[#1]{#1}
\def\refs#1{\count255=1[\r@fs #1{\hbox{}}]}
\def\r@fs#1{\ifx\und@fined#1\message{reflabel \string#1 is undefined.}%
\nref#1{need to supply reference \string#1.}\fi%
\vphantom{\hphantom{#1}}\edef\next{#1}\ifx\next\em@rk\def\next{}%
\else\ifx\next#1\ifodd\count255\relax\xref#1\count255=0\fi%
\else#1\count255=1\fi\let\next=\r@fs\fi\next}

%%%%%%%TABLE OF CONTENT%%%%%%%%%%%%%

\def\writetoc{\immediate\openout\tfile=todasym.tmp
    \def\writetoca##1{{\edef\next{\write\tfile{\noindent  ##1
    \string\leaderfill {\noexpand\number\pageno} \par}}\next}}}

%       and this lists table of contents on second pass

%
\catcode`\@=12 % at signs are no longer letters
%
%   Unpleasantness in calling in abstract and title fonts
\edef\tfontsize{\ifx\answ\bigans scaled\magstep3\else scaled\magstep4\fi}
\font\titlerm=cmr10 \tfontsize \font\titlerms=cmr7 \tfontsize
\font\titlermss=cmr5 \tfontsize \font\titlei=cmmi10 \tfontsize
\font\titleis=cmmi7 \tfontsize \font\titleiss=cmmi5 \tfontsize
\font\titlesy=cmsy10 \tfontsize \font\titlesys=cmsy7 \tfontsize
\font\titlesyss=cmsy5 \tfontsize \font\titleit=cmti10 \tfontsize
\skewchar\titlei='177 \skewchar\titleis='177 \skewchar\titleiss='177
\skewchar\titlesy='60 \skewchar\titlesys='60 \skewchar\titlesyss='60
\def\titlefont{\def\rm{\fam0\titlerm}% switch to title font
\textfont0=\titlerm \scriptfont0=\titlerms \scriptscriptfont0=\titlermss
\textfont1=\titlei \scriptfont1=\titleis \scriptscriptfont1=\titleiss
\textfont2=\titlesy \scriptfont2=\titlesys \scriptscriptfont2=\titlesyss
\textfont\itfam=\titleit
\def\it{\fam\itfam\titleit}\rm}
\font\authorfont=cmcsc10 \ifx\answ\bigans\else scaled\magstep1\fi
\ifx\answ\bigans\def\abstractfont{\tenpoint}\else \font\abssl=cmsl10 scaled
\magstep1 \font\absrm=cmr10 scaled\magstep1 \font\absrms=cmr7
scaled\magstep1 \font\absrmss=cmr5 scaled\magstep1 \font\absi=cmmi10
scaled\magstep1 \font\absis=cmmi7 scaled\magstep1 \font\absiss=cmmi5
scaled\magstep1 \font\abssy=cmsy10 scaled\magstep1 \font\abssys=cmsy7
scaled\magstep1 \font\abssyss=cmsy5 scaled\magstep1 \font\absbf=cmbx10
scaled\magstep1 \skewchar\absi='177 \skewchar\absis='177
\skewchar\absiss='177 \skewchar\abssy='60 \skewchar\abssys='60
\skewchar\abssyss='60
\def\abstractfont{\def\rm{\fam0\absrm}% switch to abstract font
\textfont0=\absrm \scriptfont0=\absrms \scriptscriptfont0=\absrmss
\textfont1=\absi \scriptfont1=\absis \scriptscriptfont1=\absiss
\textfont2=\abssy \scriptfont2=\abssys \scriptscriptfont2=\abssyss
\textfont\itfam=\bigit \def\it{\fam\itfam\bigit}\def\footnotefont{\tenpoint}%
\textfont\slfam=\abssl \def\sl{\fam\slfam\abssl}%
\textfont\bffam=\absbf \def\bf{\fam\bffam\absbf}\rm}\fi
\def\tenpoint{\def\rm{\fam0\tenrm}% switch back to 10-point type
\textfont0=\tenrm \scriptfont0=\sevenrm \scriptscriptfont0=\fiverm
\textfont1=\teni  \scriptfont1=\seveni  \scriptscriptfont1=\fivei
\textfont2=\tensy \scriptfont2=\sevensy \scriptscriptfont2=\fivesy
\textfont\itfam=\tenit \def\it{\fam\itfam\tenit}\def\footnotefont{\ninepoint}%
\textfont\bffam=\tenbf
\def\bf{\fam\bffam\tenbf}\def\sl{\fam\slfam\tensl}\rm}
\font\ninerm=cmr9 \font\sixrm=cmr6 \font\ninei=cmmi9 \font\sixi=cmmi6
\font\ninesy=cmsy9 \font\sixsy=cmsy6 \font\ninebf=cmbx9 \font\nineit=cmti9
\font\ninesl=cmsl9 \skewchar\ninei='177 \skewchar\sixi='177
\skewchar\ninesy='60 \skewchar\sixsy='60
\def\ninepoint{\def\rm{\fam0\ninerm}% switch to footnote font
\textfont0=\ninerm \scriptfont0=\sixrm \scriptscriptfont0=\fiverm
\textfont1=\ninei \scriptfont1=\sixi \scriptscriptfont1=\fivei
\textfont2=\ninesy \scriptfont2=\sixsy \scriptscriptfont2=\fivesy
\textfont\itfam=\ninei \def\it{\fam\itfam\nineit}\def\sl{\fam\slfam\ninesl}%
\textfont\bffam=\ninebf \def\bf{\fam\bffam\ninebf}\rm}
%
%---------------------------------------------------------------------
%

\hyphenation{anom-aly anom-alies coun-ter-term coun-ter-terms}
\def\inv{^{\raise.15ex\hbox{${\scriptscriptstyle -}$}\kern-.05em 1}}

\def\Dsl{\,\raise.15ex\hbox{/}\mkern-13.5mu D} %this one can be subscripted
\def\dsl{\raise.15ex\hbox{/}\kern-.57em\partial}

\def\tr{{\rm tr}} 
\font\bigit=cmti10 scaled \magstep1
 %pound sterling
\def\lspace{\ifx\answ\bigans{}\else\qquad\fi}
\def\lbspace{\ifx\answ\bigans{}\else\hskip-.2in\fi} % $$\lbspace...$$
\def\boxeqn#1{\vcenter{\vbox{\hrule\hbox{\vrule\kern3pt\vbox{\kern3pt
     \hbox{${\displaystyle #1}$}\kern3pt}\kern3pt\vrule}
    }}}
\def\mbox#1#2{\vcenter{\hrule \hbox{\vrule height#2in
         \kern#1in \vrule} \hrule}}  %e.g. \mbox{.1}{.1}

%%%%%%%%%%%%%%%%%%%%%%%%

\newwrite\ffile\global\newcount\figno \global\figno=1
\def\nfig#1{\xdef#1{fig.~\the\figno}%
\writedef{#1\leftbracket fig.\noexpand~\the\figno}%
\ifnum\figno=1\immediate\openout\ffile=figs.tmp\fi\chardef\wfile=\ffile%
\immediate\write\ffile{\noexpand\medskip\noexpand\item{Fig.\ \the\figno. }
\reflabeL{#1\hskip.55in}\pctsign}\global\advance\figno by1\findarg}
\def\xfig{\expandafter\xf@g}
\def\xf@g fig.\penalty\@M\ {}
\def\figs#1{figs.~\f@gs #1{\hbox{}}}
\def\f@gs#1{\edef\next{#1}\ifx\next\em@rk\def\next{}\else
\ifx\next#1\xfig #1\else#1\fi\let\next=\f@gs\fi\next}
\newwrite\lfile
{\escapechar-1\xdef\pctsign{\string\%}\xdef\leftbracket{\string\{}
\xdef\rightbracket{\string\}}\xdef\numbersign{\string\#}}

\def\writestop{\def\writestoppt{\immediate\write\lfile{\string\pageno%
\the\pageno\string\startrefs\leftbracket\the\refno\rightbracket%
\string\def\string\secsym\leftbracket\secsym\rightbracket%
\string\secno\the\secno\string\meqno\the\meqno}\immediate\closeout\lfile}}
\def\writestoppt{}\def\writedef#1{}
\def\seclab#1{\xdef #1{\the\secno}\writedef{#1\leftbracket#1}\wrlabeL{#1=#1}}
\def\subseclab#1{\xdef #1{\secsym\the\subsecno}%
\writedef{#1\leftbracket#1}\wrlabeL{#1=#1}}
\newwrite\tfile \def\writetoca#1{}
\def\leaderfill{\leaders\hbox to 1em{\hss.\hss}\hfill}

%%%%%% Tildes and hats%%%%%%%%%%%%%%%

\def\tilde{\widetilde}
\def\bar{\overline}
\def\hat{\widehat}

%%%%%%%%%%%%% Cech %%%%%%%%%%%%
\def\cech{${\rm C}^{\kern-6pt \vbox{\hbox{$\scriptscriptstyle\vee$}\kern2.5pt}}${\rm ech}}
\def\Cech{${\sl C}^{\kern-6pt \vbox{\hbox{$\scriptscriptstyle\vee$}\kern2.5pt}}${\sl ech}}

%%%%%%%%%%%% Greek %%%%%%%%%%%%

\def\d{{\delta}}
\def\g{{\gamma}}

\def\e{{\epsilon}}

\def\ve{{\varepsilon}}

\def\m{{\mu}}

\def\u{{\Upsilon}}
\def\l{{\lambda}}
\def\s{{\sigma}}

%%%%%%%% Calligraphic letters  %%%%%%%%%%%%%

\def\CC{{\cal C}}
\def\CD{{\cal D}}
\def\CE{{\cal E}}
\def\CF{{\cal F}}

\def\CL{{\cal L}}
\def\CM{{\cal M}}
\def\CN{{\cal N}}

\def\CP{{\cal P}}

\def\CS{{\cal S}}

\def\CV{{\cal V}}

%%%%%%%%%%% bold %%%%%%%%%%%%%%

\def\bC{{\bf C}}

\def\bF{{\bf F}}

\def\bI{{\bf I}}

\def\bL{{\bf L}}

\def\bm{{\bf m}}

\def\bP{{\bf P}}

\def\bR{{\bf R}}
\def\bS{{\bf S}}

\def\bt{{\bf t}}

\def\bW{{\bf W}}

\def\bZ{{\bf Z}}

%%%%%%%%%%%% Derivatives  %%%%%%%%%%%
\def\p{\partial}
\def\pb{\bar{\partial}}

\def\dd{{\rm d}}
%%%%%%%%%%% letters with bar %%%%%%%%

%%%%%%%%% underlined letters %%%%%%%%%%%

%
%---------------------------------------------------------------------
%

\hyphenation{anom-aly anom-alies coun-ter-term coun-ter-terms}
\def\inv{^{\raise.15ex\hbox{${\scriptscriptstyle -}$}\kern-.05em 1}}

\def\Dsl{\,\raise.15ex\hbox{/}\mkern-13.5mu D}
%this one can be subscripted
\def\dsl{\raise.15ex\hbox{/}\kern-.57em\partial}

\def\tr{{\rm tr}} 
\font\bigit=cmti10 scaled \magstep1

 %pound sterling
\def\lspace{\ifx\answ\bigans{}\else\qquad\fi}
\def\lbspace{\ifx\answ\bigans{}\else\hskip-.2in\fi} % $$\lbspace...$$
\def\boxeqn#1{\vcenter{\vbox{\hrule\hbox{\vrule\kern3pt\vbox{\kern3pt
      \hbox{${\displaystyle #1}$}\kern3pt}\kern3pt\vrule}\hrule}}}
\def\mbox#1#2{\vcenter{\hrule \hbox{\vrule height#2in
          \kern#1in \vrule} \hrule}}  %e.g. \mbox{.1}{.1}
%   matters of taste
%\def\tilde{\widetilde} \def\bar{\overline} \def\hat{\widehat}
%
% some sample definitions

\def\log{{\rm log}}

\def\darr#1{\raise1.5ex\hbox{$\leftrightarrow$}\mkern-16.5mu #1}
 %pound sterling
\def\ha{{1\over 2}}
\def\half{{\textstyle{1\over2}}} %puts a small half in a displayed eqn
\def\roughly#1{\raise.3ex\hbox{$#1$\kern-.75em\lower1ex\hbox{$\sim$}}}

\def\np#1#2#3{Nucl. Phys. {\bf B#1} (#2) #3}
\def\plb#1#2#3{Phys. Lett. {\bf #1B} (#2) #3}
\def\pla#1#2#3{Phys. Lett. {\bf #1A} (#2) #3}

\def\cmp#1#2#3{Comm. Math. Phys. {\bf #1} (#2) #3}

\def\atmp#1#2#3{Adv.~Theor.~Math.~Phys.{\bf #1} (#2) #3}

%%%%%%%%%%%%%%%  Rublenye bukvy   %%%%%%%%%%%%%%%%%
\def\IB{\relax\hbox{$\inbar\kern-.3em{\rm B}$}}

\def\ID{\relax\hbox{$\inbar\kern-.3em{\rm D}$}}
\def\IE{\relax\hbox{$\inbar\kern-.3em{\rm E}$}}
\def\IF{\relax\hbox{$\inbar\kern-.3em{\rm F}$}}
\def\IG{\relax\hbox{$\inbar\kern-.3em{\rm G}$}}
\def\IGa{\relax\hbox{${\rm I}\kern-.18em\Gamma$}}
\def\IH{\relax{\rm I\kern-.18em H}}
\def\IK{\relax{\rm I\kern-.18em K}}
\def\IL{\relax{\rm I\kern-.18em L}}
\def\IP{\relax{\rm I\kern-.18em P}}
\def\II{\relax{\rm I\kern-.18em I}}

\def\ndt{{\noindent}}

%%%%%%%% Calligraphic letters  %%%%%%%%%%%%%

\def\CC{{\cal C}}
\def\CD{{\cal D}}
\def\CE{{\cal E}}
\def\CF{{\cal F}}

\def\CL{{\cal L}}
\def\CM{{\cal M}}
\def\CN{{\cal N}}

\def\CP{{\cal P}}

\def\CS{{\cal S}}

\def\CV{{\cal V}}

\def\bC{{\bf C}}

\def\bF{{\bf F}}

\def\bI{{\bf I}}

\def\bL{{\bf L}}

\def\bP{{\bf P}}

\def\bR{{\bf R}}
\def\bS{{\bf S}}

\def\bW{{\bf W}}

\def\bZ{{\bf Z}}

%%%%%%%%%%%% Derivatives  %%%%%%%%%%%
\def\p{\partial}
\def\pb{\bar{\partial}}

\def\dd{{\rm d}}
%%%%%%%%%%% letters with bar %%%%%%%%

%%%%%%%%%% Math symbols %%%%%%%%%%%%%

\def\ch{{\rm ch}}

%%%%%%%%%%%%%% Lie algebras %%%%%%%%%%%%%%%%%%%%%%

\def\inbar{\,\vrule height1.5ex width.4pt depth0pt}

%%%%%%%%%%% Macros for boxes %%%%%%%%%%%
\def\boxit#1{\vbox{\hrule\hbox{\vrule\kern8pt
\vbox{\hbox{\kern8pt}\hbox{\vbox{#1}}\hbox{\kern8pt}}
\kern8pt\vrule}\hrule}}
\def\mathboxit#1{\vbox{\hrule\hbox{\vrule\kern8pt\vbox{\kern8pt
\hbox{$\displaystyle #1$}\kern8pt}\kern8pt\vrule}\hrule}}
\def\sqx{\mbox{.04}{.04}}
%%%%%%%%%%%%%%%%%%%%%%%%

\def\lime{{\rm Lim}_{\kern -16pt \vbox{\kern6pt\hbox{$\scriptstyle{\e \to 0}$}}}}

\def\naiveq{\qquad =^{\kern-12pt \vbox{\hbox{$\scriptscriptstyle{\rm naive}$}\kern5pt}} \qquad}

%%%%% REFS %%%%%%%%%%
\def\lref{\begingroup\obeylines\lr@f}
\def\lr@f#1#2{\gdef#1{\ref#1{#2}}\endgroup\unskip}

\lref\gkmmm{A.~Gorsky, I.~Krichever, A.~Marshakov, A.~Mironov, A.~Morozov,
{\it "Integrability and Seiberg-Witten exact solution", } hep-th/9505035,
\plb{355}{1995}{466}}
\lref\booksSW{
A.~Marshakov, {\it ``Seiberg-Witten theory and integrable systems,''}
World Scientific, Singapore (1999)\semi
``Integrability: the Seiberg-Witten and Whitham equations",
Eds. H.~Braden and I.~Krichever, Gordon and Breach (2000)}
\lref\gmmm{A.~Gorsky, A.~Marshakov, A.~Mironov, A.~Morozov,
{\it "RG equations from Whitham hierarchy"}, hep-th/9802007,
\np{527}{1998}{690-716}}
\lref\freckles{A.~Losev, N.~Nekrasov, S.~Shatashvili, {\it "The freckled instantons"},
hep-th/9908204, Yuri Golfand Memorial Volume,
Shifman~M.A. (ed.) "The many faces of the superworld", 453-475, World Scientific;
{\it "Freckled instantons in two and four dimensions"}, hep-th/9911099,
Class. Quant. Grav. {\bf 17}(2000) 1181-1187}

\lref\lowrank{K.~Intriligator, P.~Kraus, A.~Ryzhov, M.~Shigemori, C.~Vafa,
 {\it "On low rank classical groups in string theory, gauge theory and matrix models"},  hep-th/0311181 ,
\np{682}{2004}{45-82}}

\lref\bmodel{E.~Witten, {\it ``Mirror manifolds and topological field theory'',}
hep-th/9112056\semi
M.~Kontsevich, {\it ''Homological algebra of mirror symmetry''}, alg-geom/9411018}

\lref\arnold{V.I.~Arnol'd, {\it ``Mathematical methods of classical mechanics"}, Moscow,
Nauka, 1989}

\lref\ihiggs{G.~Moore, N.~Nekrasov, S.~Shatashvili, {\it "Integration over the
Higgs branches"}, hep-th/9712241, \cmp{209}{2000}{97-121} }
\lref\niklec{N.~Nekrasov, {\it Lectures on super-Yang-Mills, partitions, Mayer
expansion, and limit shapes}, UvA, May 2005}

\lref\niksch{N.~Nekrasov, A.~Schwarz, hep-th/9802068}

\lref\WittenYM{
E.~Witten, "An interpretation of classical Yang-Mills theory", Phys.Lett.B77:394,1978}

\lref\WittenCS{
E.~Witten, Chern-Simons theory as a string theory, Prog.Math.133:637-678,1995 , hep-th/9207094}

\lref\WittenZT{
E~Witten, ''Two dimensional models with (0,2) supersymmetry: perturbative aspects'', hep-th/0504078}

\lref\Blau{
M.~Blau and G.~Thompson, "Aspects of $N(T)\geq   2$ topological gauge theories and D-branes,''
Nucl.\ Phys.\ B {\bf 492} (1997) 545
[arXiv:hep-th/9612143].}

%%%%%%%% Logan %%%%%%%%%%%%

\lref\logshep{B.F.~Logan, L.A.~Shepp, {\it ``A variational problem for random Young
tableaux''},   Advances in Math.  26  (1977),  no. 2, 206--222}

\lref\kerovi{S.V.~Kerov, {\it `` Interlacing measures''},  {\sl  Kirillov's seminar on
representation theory},  35--83, Amer. Math. Soc. Transl. Ser. 2, 181,
Amer. Math. Soc., Providence, RI, 1998;
{\it ``Anisotropic Young diagrams and symmetric Jack
functions''},
(Russian)  {\cyr Funk. Anal. i Prilozhen.}  34  (2000),  no. 1, 51--64,
96;  translation in  Funct. Anal. Appl.  34  (2000),  no. 1, 41--51;
{\it ``Random Young
tableaux''}, {\cyr Teor. veroyat. i ee primeneniya}, {\bf 3} (1986), 627-628
(in Russian)\semi
A.~M.~Vershik, {\it ``Hook formulae and related identities''}, {\cyr
Zapiski sem. LOMI}, {\bf 172} (1989), 3-20 (in Russian)}

%%%%%%%%%%

\lref\abcd{N.~Nekrasov and S.~Schadchine, {\it ABCD of instantons},
\cmp{252}{2004}{359-391}, hep-th/0404225.}

\lref\atiyahbott{M.~Atiyah, R.~Bott, {\it The moment map and
equivariant cohomology} , Topology  {\bf  23}, vol. 1(1984) 1-28}

\lref\kliu{K.~Liu, {\it Holomorphic equivariant cohomology},
Math. Annalen, {\bf 303} 1 ( 1995) 125-148}

\lref\ztwitten{E~Witten, {\it Two dimensional models with (0,2)
supersymmetry: perturbative aspects}, hep-th/0504078}

\lref\griffitsharris{P.~Griffiths, J.~Harris, {\it Principles of
algebraic geometry}, 1978, New York, Wiley \& Sons}

\lref\knizhnik{V.~Knizhnik, {\it Analytic fields on Riemann surfaces},
\plb{180}{1986}{247} \semi
{\it Analytic fields on Riemann surfaces II}, \cmp{112}{1987}{567-590}}
\lref\fln{E.~Frenkel, A.~Losev, N.~Nekrasov, to appear}
\lref\aib{E.~Frenkel, A.~Losev, {\it Mirror symmetry in two steps: A-I-B},
hep-th/0505131}
\lref\swnc{N.~Seiberg, E.~Witten, {\it String theory and noncommutative geometry},
JHEP 09 (1999) 032}

\lref\wittentdg{E.~Witten, {\it Two dimensional gauge theories revisited}, hep-th/9204083}

\lref\sadov{M.Bershadsky, V.Sadov, C.Vafa}

\lref\polyakovbook{A.~Polyakov, "Gauge fields and strings", Harwood Academic Press}

\lref\niksw{N.~Nekrasov, {\it ``Seiberg-Witten prepotential from instanton counting''},
hep-th/0206161, hep-th/0306211, \atmp{7}{2004}{831-864} }
\lref\lmn{A.~Losev, A.~Marshakov, N.~Nekrasov, {\it ``Small instantons, little
strings, and free fermions''}, hep-th/0302191, Ian Kogan memorial volume, M.Shifman,
A.Vainshtein and J. Wheater (eds.) ``From fields to strings:
   circumnavigating theoretical physics", 581-621}
\lref\nikand{N.~Nekrasov, A.~Okounkov, {\it ``Seiberg-Witten theory and random
partitions''}, hep-th/0306238 }
\lref\uchebnik{I.~Macdonald, {\it ``Symmetric functions and Hall polynomials''},
Clarendon Press, Oxford, 1979}

\lref\issues{A.~Losev, N.~Nekrasov, S.~Shatashvili, {\it ''Issues in topological
gauge theory''}, hep-th/9711108 , \np{534}{1998}{549-611}}

\lref\egu{T.~Eguchi, S.-K.~Yang, {\it ``The topological ${\bC\bP}^1$ model and the
large-$N$ matrix integral"}, Mod.~Phys.~Lett. {\bf A}9 (1994) 2893-2902, hep-th/9407134}
\lref\tks{
 T.~Eguchi, K.~Hori, S.-K.~Yang,
{\it ''Topological $\sigma$-models and large-$N$ matrix integral''}, Int.~J.~Mod.~Phys.
{\bf A}10 (1995) 4203, hep-th/9503017}

\lref\kriw{I.~Krichever, {\it "The $\tau$-function of the universal Whitham hierarchy,
matrix models and topological field theories"}, hep-th/9205110,
Commun. Pure. Appl. Math. {\bf 47} (1992) 437.}
\lref\pqdual{S.~Kharchev and A.~Marshakov,
{\it "Topological versus non-topological theories and $p-q$ duality in
$c \le 1$ 2d gravity models}, hep-th/9210072, in "String theory,
quantum gravity and the unification of the fundamental interactions",
M.Bianchi, F.Fucito, E.Marinari, A.Sagnotti (eds.),
World Scientific, 1993, 331-346; {\it Integral representations and
$p-q$ duality in $c \leq  1$ 2d gravity}, hep-th/9303100,
Int. J. Mod. Phys. {\bf A10} (1995) 1219.}
\lref\taktak{K.~Takasaki, T.~Takebe, {\it "Quasi-classical limit of
Toda hierarchy and W-infinity symmetries"}, hep-th/9301070,
Lett.~Math.~Phys. 28 (1993) 165-176;
{\it ``Integrable hierarchies and dispersionless limit"}, hep-th/9405096,
Rev.Math.Phys. {\bf 7} (1995) 743-808.}
\lref\comamo{A.~Marshakov, A.~Mironov, A.~Morozov,
  {\it ``Generalized matrix models as conformal field theories: Discrete case"},
  Phys.\ Lett.\  {\bf B265}, 99 (1991)\semi
  S. Kharchev, A. Marshakov, A. Mironov, A. Morozov, S. Pakuliak,
{\it ``Conformal matrix models as an alternative to conventional multi-matrix models"},
hep-th/9208044, Nucl.Phys. {\bf B404} (1993) 717-750.}
\lref\gecomamo{G.Akemann,
{\it ``Higher genus correlators for the Hermitian matrix model
with multiple cuts"}, hep-th/9606004, Nucl.Phys. {\bf B482} (1996) 403\semi
I.~K.~Kostov,
{\it ``Conformal field theory techniques in random matrix models"}, hep-th/9907060\semi
R.Dijkgraaf, A.Sinkovics, M.Tem\"urhan,
{\it ``Matrix models and gravitational corrections"},
hep-th/0211241.}
\lref\amv{A.Marshakov, {\it ``Strings, integrable systems, geometry and
statistical models"}, hep-th/0401199.}

\def\frac#1#2{{{#1}\over{#2}}}

\Title{
\vbox{\baselineskip 10pt
\hbox{FIAN/TD-10/06}
\hbox{ITEP-TH-57/06}
\hbox{IHES-P/06/43}
}}{
\vbox{\vskip 30 true pt
\smallskip
\centerline{EXTENDED SEIBERG-WITTEN THEORY}
  \bigskip \centerline{AND INTEGRABLE HIERARCHY} \vskip2pt}}

\bigskip
\centerline{Andrei Marshakov$^{1}$, Nikita A. Nekrasov\footnote{$^{\star}$}{On
leave of
absence from: ITEP, Moscow, Russia}$^{2}$}
\bigskip\bigskip
\centerline{\it $^1$ Lebedev Physics Institute and ITEP, Moscow, 117218 Russia}
\centerline{\it $^{2}$ Institut des Hautes Etudes Scientifiques,
91440 Bures-sur-Yvette, France\footnote{$^{\star\star}$}{permanent address}}
\centerline{\it $^{2}$ Department of Physics, Rutgers University, Piscataway,
NJ 08544 USA}
\bigskip \bigskip
\ndt
The prepotential of the effective ${\CN}=2$ super-Yang-Mills theory, perturbed in
the ultraviolet by the descendents $ \int d^{4}{\theta} \, {\tr} \, {\bf\Phi}^{k+1}$
of the single-trace chiral operators, is
shown to be a particular tau-function of the quasiclassical Toda hierarchy. In
the case of noncommutative $U(1)$ theory (or $U(N)$ theory with $2N-2$
fundamental hypermultiplets
at the appropriate locus of the moduli space of vacua) or a theory on a
single fractional $D3$ brane at the $ADE$ singularity the hierarchy is the
dispersionless Toda chain, and we present its explicit solution. Our results
generalise the limit shape analysis of Logan-Schepp and Vershik-Kerov, support
the prior work \lmn, which established the equivalence of these ${\CN}=2$
theories with the topological $\bf A$ string on ${\bC\bP}^1$, and clarify the
origin of the
Eguchi-Yang matrix integral. In the higher rank case we find an appropriate
variant of the quasiclassical tau-function, show how the Seiberg-Witten curve
is deformed by Toda flows, and fix the contact term ambiguity.

\vfill\eject
\centerline{\authorfont TABLE OF CONTENTS}\nobreak
{\bf     \medskip{\baselineskip=12pt\parskip=0pt\input todasym.tmp \bigbreak\bigskip}}

\newsec{Introduction}

The supersymmetric ${\CN}=2$ theories in two and four dimensions are quite
interesting both
for physicists and mathematicians. Their low-energy dynamics is rich,
yet the supersymmetry is large enough to impose rather stringent constraints
on the effective action.

\vskip 15pt
\ndt
{\it Prepotential of effective theory}

\ndt
In quantum field theory one is interested in the low-energy effective action.
The extended supersymmetry  implies that all the terms with at most two derivatives
and at most four fermions in the effective action are related,
and can be expressed through derivatives of a single holomorphic quantity, a
locally defined function ${\CF} ({\vec a})$ on the moduli space $\CM$ of vacua,
the so-called prepotential.

In the recent years the exact calculations of the prepotential were performed,
which utilized the fact that only the one-loop perturbative
corrections about (an arbitrary) instanton solution summed over the
instanton sectors, contribute to the prepotential. Actually, the prepotential
itself is not so easy to calculate. However, it turns out that a certain
"quantum corrected" prepotential
\eqn\qcorr{Z ({\vec a}, {\hbar}) = {\exp} \
\sum_{g=0}^{\infty} F_{g}({\vec a})\ {\hbar}^{2g-2}}
has a simple expression in terms of the gauge theory instantons.
The leading term ${\CF}({\vec a}) = F_{0}({\vec a})$ is the prepotential of
the low-energy effective theory. The higher order terms, $F_{g}({\vec a})$,
describe the coupling of the theory to the ${\CN}=2$ supergravity multiplet.
In the string theory realisations of the ${\CN}=2$ theory the terms
$F_{g}({\vec a})$ are computed by the genus $g$ string amplitudes.

The prepotential $F_{0}$ and the higher genus corrections $F_g$
are special in the sense that they determine the terms in the Lagrangian of
effective theory, given by the integrals over half of the superspace, the so-called
generalised F-terms.
Formally one can consider the similar F-terms in the ultraviolet theory.
The original microscopic Lagrangian
\eqn\lorg{
L_{0} = \ha\int d^{4}{\theta}\  {\tau}_{0}{\tr} {\bf\Phi}^{2} +
\ha\int d^{4}{\bar\theta} \ {\bar\tau}_{0}{\tr} {\bar{\bf\Phi}}^{2}}
can be deformed by adding the operators
\eqn\ldef{
L_{\bt} = L_{0} +  \sum_{k>0}{t_{k} \over k+1} \int d^{4}{\theta} \ {\tr}
{\bf\Phi}^{k+1} = L_{0} +  \int\ d^4{\theta} \ {\bt} \left( {\bf\Phi} \right)
}
where
\eqn\phis{{\bf\Phi} = {\Phi} + {\theta} {\psi} + {\theta}{\theta} F^{-} + \ldots }
is the vector superfield, and it is often convenient to work
with the generating function
\eqn\btf{{\bt}(x) = \sum_{k>0} t_{k} {x^{k+1} \over k+1}
}
The bare couplings ${\tau}_{0}$, ${\bar\tau}_{0}$ are adjusted so as
to produce the finite couplings in the effective Lagrangian, obtained by integrating out the high frequency modes.

\vskip 15pt
\ndt
{\it Renormalisation group flow}

\ndt
We can reformulate our problem above as the calculation of the effect of the Wilsonian
renormalisation group flow on the prepotential of the theory. One
starts with the theory with ${\CN}=2$ supersymmetry, which is determined by the
ultraviolet prepotential ${\CF}_{\rm UV}$, perturbed by arbitrary powers of the
holomorphic operators
\eqn\fuv{
{\CF}_{\rm  UV} = {\half} ( {\tau}_{0} +  t_{1})\  {\tr}\ {\Phi}^{2} +
\sum_{k > 0} t_{k}\ {{\tr}\ {\Phi}^{k+1}\over k+1}
}
and quadratic ${\overline{\CF}_{\rm UV}}$, i.e.
$$
{\overline{\CF}_{\rm UV}} = {\half}{\bar\tau}_{0}\ {\tr}\ {\bar \Phi}^{2}
\ .
$$
Then one integrates out the fast modes, i.e. the perturbative
fluctuations with momenta above certain scale ${\mu}$ as well as the non-perturbative modes, e.g. instantons
(and fluctuations around them) of all sizes smaller
then ${\mu}^{-1}$. The resulting effective theory has a derivative expansion
in the powers ${\p}^{2}\over {\mu}^{2}$. The leading terms in the expansion are all determined,
thanks to the ${\CN}=2$ supersymmetry,
by the effective prepotential ${\CF}({\mu})$. As ${\mu}$ is lowered all the
way down to zero, we arrive at the infrared prepotenial ${\CF}_{\rm IR}$:
$$
{\CF}_{\rm UV} \longrightarrow {\CF}_{\rm IR}
$$
The supersymmetry considerations suggest that the renormalisation flows of
${\CF}$ and ${\bar\CF}$ proceed more or less independently from each other.
Thus one can simplify the problem by taking the limit, ${\bar\tau}_{0} \to i \infty$,
while ${\CF}_{\rm UV}$ kept fixed. In this limit the
path integral is dominated by the gauge instantons. The setup of \niksw\ allows to
evaluate their contribution, as well as the contribution of the fluctuations around
the instantons, exactly. The price one pays is the introduction of extra parameters
into the problem, some sort of the infrared cutoff, which we denote
by ${\hbar}^{-2}$, since it appears to be a parameter of the loop expansion
in dual topological string theory \lmn. As we send $\hbar \to 0$,
the infrared cutoff is removed, and the prepotential is recovered as the
extensive part of the free energy, cf. \qcorr. For details the reader is
invited to consult \niksw\lmn\nikand.

\vskip 15pt
\ndt
{\it Prepotential and symplectic geometry}

\ndt
The goal of this paper is to extract the generalised prepotential,
${\CF}_{\rm IR}={\CF}({\vec a}, {\bt})$
as a function of the moduli of the vacua $\vec a$, and the higher Casimir couplings
$\bt$. It has been proposed in \gkmmm\ (see also \booksSW\ and references therein)
that the answer is given by tau-function of a quasiclassical or universal Whitham
integrable hierarchy \kriw\ and below we derive this hierarchy directly from
the results of instanton calculus.

For fixed ${\bt}$ the prepotential defines a generating function of a
Lagrangian submanifold ${\CL}_{\bt}$ in the complex symplectic vector space
${\bC}^{N}$,  invariant under the action of a certain  discrete
subgroup $\Gamma$ of ${\rm Sp}(2N, {\bZ})$, the group of electric-magnetic
dualities
\issues. The deformations of such submanifolds are constrained by the
considerations of duality, however at each order of deformation beyond
the first one there are ambiguities:
\eqn\defprep{\eqalign{
& {\CF} ( {\vec a}, {\bt} ) = {\CF} ( {\vec a}, 0) +
\sum_{k>0} t_{k} u_{k+1}  \cr
& \qquad\qquad +
\sum_{k, l >0} t_{k}t_{l}\left(- {{\p u_{k+1}}\over{{\p a_{n}}}}
{{\p u_{l+1}}\over{\p a_{m}}}  {\p \over  \p{\tau}_{mn}}
{\rm log}\ {\vartheta}_{E} \left( 0 \vert {\tau} \right) +
{\CC}_{kl} (u) \right) + \ldots  \cr}
}
where $u_{k} \sim \langle {\tr} {\Phi}^{k} \rangle$ are
$\Gamma$-invariant functions on ${\CL}_{0}$, i.e. the polynomials
of the coefficients of Seiberg-Witten curve,   ${\CC}_{kl}(u)$
are some polynomials of $u$, called Losev-Shatashvili polynomials, and the precise definition of the
theta constant and the theta characteristics in \defprep\ is at
the moment immaterial. While the choice of $u_{k}$ is more or less
a matter of convention, the choice of Losev-Shatashvili polynomials ${\CC}_{kl}(u)$ is physically
important.  The appearance of theta functions in \defprep\ is dictated by the
$\Gamma$-invariance and the extra terms ${\CC}_{kl}(u)$ are
constrained by the degree considerations \issues, which predict
that for small $k,l$, they vanish.
However, to determine them
%for all $m,n$
one needs a microscopic theory, and
below we show how a particular choice of microscopic theory
determines all contact terms ${\CC}_{kl}(u)$ in terms of quasiclassical
integrable hierarchy. In particular, we shall see that
Losev-Shatashvili polynomials ${\CC}_{kl}(u)$ indeed vanish for $k,l<N$, i.e.
exactly for the Casimirs which deform the Seiberg-Witten geometry for the
$U(N)$ gauge theory. This distinguishes the quasiclassical hierarchy
derived in this paper as directly coming from the microscopic instanton theory,
compared to the previous attempts, proposed in \gmmm, based on the analogies with Landau-Ginzburg models.

\vskip 15pt
\ndt
{\it Ensembles of partitions}

\ndt
In particular, for the so-called noncommutative $U(1)$ theory (which supports instantons, \niksch), or the theory
on a single D3 brane in the background, which preserves only sixteen supercharges (so that
the theory on the brane has only eight supercharges),
the instanton partition function $Z(a, {\hbar}, {\bt})$,
${\bt} = \left( t_{1} , t_{2} , \ldots \right)$, can be shown to be given by
the sum over the Young diagrams, i.e. over the partitions \niksw\lmn\nikand:
\eqn\zuone{Z ( a, {\bt}, {\hbar} ) =
\sum_{{\l}}
{{\bm}_{\l}^2\over{( - {\hbar}^2)^{| {\l} |}}} \
{\exp} \ {1\over {\hbar}^{2}} \sum_{k>0}
t_{k}{{\ch}_{k+1} ( a, {\l})\over k+1}
}
(we shall remind the relevant notions of the theory of partitions in the Appendix). This
theory can also be realised at a special point on the moduli space of $U(N)$ gauge
theory with $2N-2$ fundamental hypermultiplets\footnote{$^{\dagger}$}{In a certain
sense, all the instanton contributions in the "U(1)" theory are the artefacts of the
imbedding of the theory into the theory with rich ultraviolet structure. In the terminology
of \freckles\ these are "freckled instantons" and their contribution must be subtracted by
the appropriate
"mirror map".  See also \ref\tim{T.~Hollowood, {\it "Calculating the prepotential
by localization on the moduli space of instantons"}, hep-th/0201075;
{\it "Testing Seiberg-Witten theory to all orders in the instanton expansion"}, hep-th/0202197} for the related and more
detailed discussion, and \lowrank\ for the related ${\CN}=1$ considerations.}.
If the theory has the gauge group $U(N)$, e.g. it is realised on the
stack of $N$ fractional D3 branes, the corresponding partition function is given by
the generalisation of \zuone:
\eqn\zunntre{Z ( {\vec a},  {\bt}, {\hbar} ) =  Z^{\rm pert} ( {\vec a} , {\bt}, {\hbar}  )
\sum_{{\vec\l}} \left( { {\bm}
( {\vec a}, {\vec \l}, {\hbar})  } \right)^{2} \ (-1)^{| {\vec\l}|}
{\exp} \ {1\over {\hbar}^{2}} \sum_{k>0} t_{k}
{{\ch}_{k+1}( {\vec a}, {\vec\l})\over k+1}
}
where ${\bm}( {\vec a}, {\vec \l}, {\hbar})$ is the $U(N)$ generalisation of
Plancherel measure \niksw\nikand\ and $ Z^{\rm pert} ( {\vec a} , {\bt}, {\hbar}  ) $
is the perturbative partition function.

%%%%%%%%%%%%%%%%%% Combinatorics etc %%%%%%%%%%%%%%%%%
\lref\verker{ S.~V.~Kerov,
A.~M.~Vershik, {\it ``Asymptotics of the Plancherel measure of the symmetric
group and the limit shape of the Young diagrams''}, {\cyr DAN SSSR}, {\bf
233} (1977), no. 6,  1024-1027 (in Russian)}
\lref\verkeri{S.~V.~Kerov,
A.~M.~Vershik, {\it
`` Asymptotic behaviour of the maximum and
generic dimensions of irreducible representations of the symmetric
group,''} (Russian)  {\cyr Funk. Anal. i Prilozhen.}  {\bf 19}  (1985),  no. 1,
25--36}
\lref\kerovii{
S.V.~Kerov, {\it ``Gaussian limit for the Plancherel measure of the
symmetric group''},  C. R. Acad. Sci. Paris S'r. I Math.  {\bf 316}  (1993),  no.
4, 303--308.}

%%%%%%%%%%%%

\vskip 15pt
\ndt
{\it Limit shape and Toda chain}

\ndt
We shall evaluate  the sums like \zuone\ by the analogue of the saddle
point method. The limit ${\hbar} \to 0$ of the sum \zuone\ is dominated
by a partition ${\l}_{*}$,
of a large size $\sim {\hbar}^{-2}$. The shape of the Young diagram of
this partition, the so-called limit shape, is found by extremizing the
effective energy functional.
When all but the first coupling $t_{1}$ are set to zero, this limit shape
is the celebrated "arcsin law" curve, found by Vershik-Kerov and
Logan-Shepp \logshep\verker\verkeri\kerovii.
Our main claim for the rank $N=1$ is that the evolution of the
limit shape under the higher
Casimirs is governed by the quasiclassical Toda hierarchy, and we
present the corresponding solution explicitly.
This result, which we prove in the section $\underline{3}$, gives a further
confirmation of the claim of the previous work \lmn\ of A.~S.~Losev and the authors,
that the BPS sector of the rank one ${\CN}=2$
gauge theory is equivalent to the stationary sector of the ${\bC\bP}^{1}$
Gromov-Witten theory. The dispersionless Toda hierarchy is well-known to
describe the genus zero part of that theory (see, e.g. \egu).

In the section $\underline{4}$ we describe the non-abelian theory. We find that the
corresponding limit shape is described by the Krichever quasiclassical
tau-function \kriw\
associated with a family of hyperelliptic curves with two marked points
\eqn\hypell{
y^{2} = \prod_{l=1}^{N} ( z - x_{l}^{+}({\vec a}, {\bt}) )
( z - x_{l}^{-}( {\vec a}, {\bt}))
}
Interestingly enough, for nonvanishing ${\bf t}$
this family extends outside the family of Seiberg-Witten curves.
The answer is encoded in \hypell\
and a particular $(1,0)$-differential $d\Sigma$,
$$
d{\Sigma} = {{\sigma}(z; {\vec a}, {\bt} ) dz\over y} \ ,
$$
holomorphic outside the points $P_{\pm}$ where $z = \infty$, which obeys
certain normalisation conditions. We demonstrate how the instanton corrections
(for nonvanishing ${\bf t}$) can be extracted from the generalised Seiberg-Witten
geometry, and derive equation \defprep\ from the "first principles". In the section
$\underline{5}$
discuss the relation of our approach with the Eguchi-Yang matrix integral and
propose a possible way to extend this relation beyond the quasiclassical theory.

\newsec{Gauge theory partition function}

We study ${\CN}=2$ gauge theory in the self-dual $\Omega$-background.
The setup and the relevant physics are reviewed in \niksw\lmn\nikand\ so
we just briefly discuss it here. The gauge theory path integral is saturated
by instantons, and ${\CN}=2$ supersymmetry cancels the contribution of
fluctuations, so that effectively one has to integrate "unities" over
the corresponding moduli spaces of the instanton solutions. These moduli spaces
can be described by the ADHM construction, and the corresponding integrals can be
computed via equivariant localisation technique in nontrivial $\Omega$-background
\niksw. As a result, the gauge theory partition function can be presented
effectively in one of the following ways:

\item{1)} The sum over instantons can be interpreted as a
Van-der-Vaals gas in one dimensions: the integrals
 over the instanton moduli space of charge $k$
 can be reduced, via ADHM construction and
 equivariant localisation a la \ihiggs\ to the grand canonical ensemble of
 a one-dimensional gas of particles
\eqn\instgas{\eqalign{ & Z ({\vec a} , {\bt}, \epsilon_{1,2} ) =
Z^{\rm pert} ({\vec a} , {\bt}, \epsilon_{1,2} )  \times \cr
\qquad & \qquad\qquad \times\sum_{k=0}^{\infty}
{1\over k!}\left({\epsilon_1+\epsilon_2\over \epsilon_1\epsilon_2}\right)^k\oint
\prod_{I=1}^k  \ d\phi_I \exp \left( -{1\over{{\e}_{1} {\e}_{2}}} U(\phi_I) \right)\times  \cr
&  \qquad\qquad\qquad\qquad\qquad\ \times\prod_{I < J}
\exp \left( -{1\over{{\e}_{1} {\e}_{2}}} V({\phi}_I- {\phi}_{J}) \right)
\cr
& {\rm where}  \ U(x)  = {\bt} ( x ) + {\bt} ( x + {\e}_{1} + {\e}_{2}) -
{\bt}( x + {\e}_{1}) - {\bt}(x + {\e}_{2}) + \cr & \qquad\qquad\qquad
+ {\e}_1{\e}_2 \sum_{l=1}^{N} {\log} \left( \left( x - a_{l} \right)^2 -
{1\over 4} \left( {\e}_{1} + {\e}_{2} \right)^2 \right)  \cr
& \qquad\qquad\qquad \ V (x) = - {\e}_{1}{\e}_{2} {\log} {x^2 ( x^2 -
\left( {\e}_{1} + {\e}_{2} \right)^2 ) \over \left( x^2  - {\e}_{1}^{2} \right)
\left( x^2 - {\e}_{2}^2 \right)} \cr}}
interacting via pair-wise Van-der-Vaals
 kind of potential, in the presence of $N$ (for $U(N)$ theory) sources. Here
 $\epsilon_{1,2}$ are parameters of the $\Omega$-background, which can be, after
 calculating integrals in \instgas\ via residues, put to be
 $\epsilon_1=-\epsilon_2=\hbar$.

\item{2)} The gauge theory partition function can be shown to be equal to the
partition function of the statistical model of random partitions.
For $U(N)$ gauge theory one deals with the ensemble of $N$-tuples of partitions
${\vec \l} = ({\l}_{1}, \ldots , {\l}_{N})$, and
the partition function of the gauge theory is \niksw\nikand:
\eqn\zunonab{Z ( {\vec a} , {\bt}, {\hbar} ) =
 Z^{\rm pert} ( {\vec a} , {\bt}, {\hbar}  )  \cdot
 \sum_{{\vec \l}} \ {\bm}^2 ( {\vec a},
 {\vec \l}, {\hbar})
 ( - 1 )^{| {\vec \l} |} {\exp} {1\over {\hbar}^{2}}\sum_{k=1}^{\infty}
 t_{k} {{\ch}_{k+1} ( {\vec a}, {\vec \l})\over k+1}
 }

\item{3)} The gauge theory partition function can be also written as
a path integral in the theory of a free chiral fermion on a Riemann sphere \lmn, or,
via bosonisation, as a path integral in the theory of a free boson ${\varphi}$
with the action:
\eqn\csacs{
{\CS} = {1\over 4{\pi}} \int_{{\bS}^{2}} {\p\varphi}{\pb\varphi}
+ {1\over\hbar}\oint_{C_{\ve}}  J w^{-1}
+ {1\over\hbar}\oint_{C_{1/{\ve}}} J w
+ \oint_{C_{1}} \sum_{k} t_{k} {\hbar}^{k-1} \left( J^{k+1} + \ldots \right)}
with the boson $\varphi$ normalised so that it takes values in a circle of
finite circumference. In \csacs\  $C_{r}$ denotes a circle $| w | = r$, and
${\ve} \downarrow 0$.
Similar path integrals were studied before in the context of
so called conformal matrix models \comamo, though with very different properties of genus
expansion \gecomamo.
In this paper we shall concentrate on the quasiclassical computations in the theory \csacs.

We start with reminding the basic facts about partitions and free fermions.

\subsec{Partition function and partitions}

{\it The partition} ${\l}$ of the size $| {\l} | = {\l}_{1} + {\l}_{2} +
\ldots + {\l}_{{\ell}_{\l}}$ is a non-increasing sequence of non-negative
integers ${\l}_{i} \in {\bZ}_{\geq 0}$ (some details and examples are collected
in Appendix A):
\eqn\lp{{\l} = \left( {\l}_{1} \geq {\l}_{2} \geq {\l}_{3} \ldots \geq
{\l}_{{\ell}_{\l}} > {\l}_{{\ell}_{\l} + 1} = {\l}_{{\ell}_{\l}+2} = \ldots = 0 \right)}
For our purposes it is
convenient to encode the partition ${\l}$ in the so-called
{\it profile function} $f_{\lambda}(x)$, which is a piece-wise linear function,
given by:
\eqn\proff{\eqalign{ f_{\lambda}(x) = \bigl| x - a\bigr| + & \  \cr
\qquad\qquad +  &\  \sum_{i=1}^{\infty}
\quad \bigl| x - a - {\hbar} ( {\l}_{i} - i +1 ) \bigr| - \bigl| x - a -{\hbar}
( {\l}_{i} - i  ) \bigr| \cr
& \qquad\qquad - \bigl| x - a - {\hbar} (1-i) \bigr|+ \bigl| x - a + {\hbar}i
 \bigr|  \cr}}
The Chern character of the
``universal sheaf'' ${\rm ch}({\CE})$ (at the fixed point in instanton
moduli space, characterized
by partition $\lambda$, see \lmn\niksw\ and references therein)
is essentially the Fourier
transform of the profile function:
\eqn\chtrpro{
{\rm ch}({\CE}) = {\half} \int \ dx \ f_{\lambda}''(x)\ e^{u x}
= e^{u a} \left( 1 + ( 1 - e^{-u {\hbar}}) \sum_{i=1}^\infty e^{u {\hbar}( 1 - i) }
( e^{u {\hbar}{\l}_{i}} - 1 ) \right)
}
hence for the coefficients of its expansion
\eqn\gauch{
 {\rm ch}({\CE}) = \left( e^{{\hbar}u\over 2} -
e^{-{\hbar u \over 2}} \right) {\rm ch}_{\l} \left( {a\over {\hbar}} , {\hbar}u  \right)
 = \sum_{k=0}^{\infty}\ {u^k\over k!}\ {\rm ch}_{k}( a, {\l})
 }
which
enter the formula for the statistical weight of the partition $\l$ in the
ensemble \zuone, one has
\eqn\chnrp{
{\rm ch}_{k}(a, {\l}) = \ha \int \ dx \ f_{\lambda}''(x) x^k
\sim \sum_{i=1}^{\infty}
\left( (a + {\hbar} ( {\l}_{i} - i +1 ))^k - (a +{\hbar}
( {\l}_{i} - i  ))^k\right)
}
In the nonabelian case, for the gauge group $U(N)$,
the universal sheaf $\CE$ splits as a sum of $N$ rank one sheaves ${\CE}_{l}$. Accordingly, at the fixed point of the torus action on the moduli space of
instantons, corresponding to the $N$-tuple of partitions ${\vec \l}$:
\eqn\gauchna{
{\rm ch}( {\CE}) = \left( e^{ {\hbar u}\over 2} - e^{-{\hbar u\over 2}} \right)
\sum_{l=1}^N e^{u a_{l}}  {\rm ch}_{{\l}^{(l)}} ( 0 , {\hbar}u) =
\sum_{k=0}^{\infty} {u^{k}\over k!}\ {\rm ch}_{k}( {\vec a}, {\vec\l})}
with
\eqn\chk{
{\rm ch}_{k}( {\vec a}, {\vec\l}) = \sum_{l=1}^{N} {\rm ch}_{k}( a_{l}, {\l}_{l})
}
For the empty partitions ${\l}^{(l)} = \emptyset$ the Chern character
\gauchna\ reduces to the generating function of the vacuum expectation
values of the single trace operators ${\tr}\, {\Phi}^{k}$, in
the absence of quantum corrections.

\vskip 15pt
\ndt
{\it Plancherel measure and the profile of the partition}

\ndt
The Plancherel measure
\eqn\planch{
{\bm}_{\l} = \prod_{i < j} {{{\l}_{i} - {\l}_{j} + j - i}\over {j-i}} =
\prod_{i=1}^{{\ell}_{\l}} {( {\ell}_{\l} - i )! \over
( {\ell}_{\l} + {\l}_{i} - i )! }
\prod_{1 \leq i < j \leq {\ell}_{\l}} {{{\l}_{i} - {\l}_{j} + j - i}\over {j-i}}}
can also be expressed in terms of the
profile function $f_{\lambda}(x)$:
\eqn\planchf{\frac{{\bm}^{2}_{\l}}{(-\hbar^{2})^{|{\l}|}} =
{\exp} \ \left( -\frac{1}{4}\int_{x_{1}>x_{2}} dx_{1} dx_{2}
\ f_{\lambda}''(x_{1})\ f_{\lambda}''(x_{2}) \
{\gamma}_{\hbar} (x_{1}-x_{2})\right)}
where
the kernel ${\gamma}_{\hbar} ( x )$ solves the following difference equation:
\eqn\diffrnk{{\gamma}_{\hbar}(x + \hbar ) + {\gamma}_{\hbar}
(x - \hbar) - 2 {\gamma}_{\hbar} (x) = {\rm log} x^2  }
and is given by the asymptotic series:
\eqn\gammser{{\gamma}_{\hbar}(x) = {1\over \hbar^2}
\left( {\bF}(x) - \frac{{\hbar}^{2}
}{12} {\bF}^{(2)}(x) + \frac{{\hbar}^4}{240} {\bF}^{(4)}(x) -
\frac{{\hbar}^{6}}{6048} {\bF}^{(6)}(x) + \ldots \right)}
determined by its leading term, or the
 function:
\eqn\SWfun{
{\bF}(x)= x^2\left(\log \ x -{3\over 2}\right)
}
which enters the perturbative prepotential of pure ${\CN}=2$ super-Yang-Mills theory.
The finite difference derivative of $\gamma_{\hbar}(x)$ is related to the
$\Gamma$-function:
\eqn\fstdffr{{\gamma}_{\hbar} \left( x + \frac{\hbar}{2}\right) -  {\gamma}_{\hbar}
\left( x - \frac{\hbar}{2}\right) = {\rm log} \left( {\hbar}^{{x\over\hbar}}
{\Gamma} \left( {x\over \hbar}  + \frac{1}{2} \right) \right)  }
In the $\hbar\to 0$ limit \diffrnk\ just turns into
${\bF}''(x)= 2\log \ x $.
The function ${\gamma}_{\hbar}(x)$ is related to the generalized Riemann zeta function:
\eqn\gfnn{{\gamma}_{\hbar}(x) = -{\zeta}^{\prime}(-1) + {1\over 12} {\rm log}{\hbar} +
{d\over ds}\Biggr\vert_{s=0} {1\over{{\Gamma}(s)}} \int_{0}^{\infty} {dt\over t} t^{s}
{e^{-t x} \over ( e^{t\hbar} - 1)(e^{-t\hbar}-1)}}
The $\zeta^{\prime}(-1)$ and ${\log}{\hbar}$ terms in \gfnn\ are chosen so as to ensure
${\g}_{\hbar}(0) = 0$. These corrections do not give any contribution to
quasiclassical part
and can be therefore simply absorbed into normalisation (in particular we did not
write them in \gammser ).

\subsec{Fermions and Baker-Akhiezer functions}

The gauge theory partition function \zuone,\zunntre\ can be compactly written
as a matrix element in the infinite wedge
representation of the group $GL({\infty})$, in other words, it
has a free fermion representation \lmn\nikand (see also
\ref\wang{W.~-P.~Li, Z~.Qin, W.~Wang, {\it "Hilbert schemes,
integrable hierarchies, and Gromov-Witten theory"},
Internat.~Math.~Res.~Notices {\bf 40} (2004), 2085-2104, math.AG/0302211}).
We shall recall it now in order to motivate the introduction of some
quasiclassical objects,
like the multi-valued functions $S(z)$ and ${\Phi}(z)$,
which will be the main tool for solving our problem. As we shall see, they  appear in the
asymptotic expansion of the fermionic one-point functions, or the Baker-Akhiezer
functions (cf. \taktak\ref\mntt{T.~Maeda, T.~Nakatsu, K.~Takasaki, T.~Tamakoshi,
{\it "Free fermion and Seiberg-Witten differential in random plane partitions"},
hep-th/0412329; {\it "Five-dimensional supersymmetric Yang-Mills
theories and random plane partitions"}, hep-th/0412327}).

\vskip 15pt
\ndt
{\it Free fermions and partitions}

\ndt
Introduce the free fermion fields, ${\psi}(w), {\tilde\psi}(w)$:
\eqn\frfrms{\eqalign{& {\psi}(w) = \sum_{r \in {\bZ} + {\half}}
{\psi}_{r} w^{-r} \left( {dw \over w} \right)^{\half} \cr
& {\tilde\psi}(w) = \sum_{r \in {\bZ} + {\half}}
{\tilde\psi}_{r} w^{r} \left( {dw \over w} \right)^{\half} }}
where$\{ {\psi}_{r} , {\tilde\psi}_{s} \} = {\d}_{rs}$.
The vacuum with the charge $M$ is defined as:
\eqn\vacm{\vert M \rangle =
\vert M ; {\emptyset} \rangle = {\psi}_{-M + \frac{1}{2}}
{\psi}_{-M + \frac{3}{2}} {\psi}_{-M + \frac{5}{2}} \ldots }
and is annihilated by the fermion harmonics:
\eqn\vacmm{\eqalign{
& {\psi}_{r} \vert M \rangle = 0, \quad r > -M, \cr
& {\tilde\psi}_{r} \vert M \rangle = 0, \quad r < -M\cr}}
We always use the normal ordering $:(\ldots):$ with respect to the
vacuum $| 0 \rangle$:
\eqn\nrmlord{\eqalign{
: {\psi}_{r} {\tilde \psi}_{s} :\quad  = \quad & \quad\ {\psi}_{r}
{\tilde \psi}_{s}\ , \quad\qquad s < 0 \cr
: {\psi}_{r} {\tilde \psi}_{s} : \quad = \quad & \ -  {\tilde \psi}_{s}
{\psi}_{r} \ , \quad\qquad r > 0 \cr}}
Let us introduce the  $W_{1+\infty}$ algebra operators:
\eqn\wgen{\eqalign{{\bW}_{k+1} = -\ & {{\hbar}^{k}\over k+1}
\oint : {\tilde\psi} \left( \left( D + {1\over 2}\right)^{k+1}  -
\left( D - {1\over 2} \right)^{k+1} \right) {\psi} : \cr
= \ & \ {{\hbar}^{k} \over k+1}
\sum_{r \in {\bZ} +{\half}}   \left[  (- r+{\half})^{k+1} -
(- r - {\half} )^{k+1} \right] \ : {\psi}_{r}{\tilde\psi}_{r} :
\cr}}
where $D = w{\p}_{w}$ and $k\geq 0$. For example,
\eqn\wgenw{
 {\bW}_{1} =  -J_{0}\ ,
 {\bW}_{2} = \hbar L_{0}\ , \quad \ldots
}
i.e. $J_{0}$ is the zero mode of the $U(1)$ current,
\eqn\uoncur{\eqalign{&  J(w) =  \ : {\tilde \psi} (w) {\psi} (w) : \ =
\sum_{k \in {\bZ}}
J_{k} w^{-k} {dw\over w} \cr  &\qquad
J_{k} = \sum_{r \in {\bZ} + {\half}} : {\tilde\psi}_{r} \psi_{r+k} : \cr}}
while the $L_{0}$ Virasoro
generator is the zero mode of the stress-energy tensor:
\eqn\virr{\eqalign{&  T(w) =  -\ :  {\tilde \psi} (w)  d{\psi} (w): \ =
\sum_{k \in {\bZ}}
L_{k} w^{-k} \left({dw\over w}\right)^2 \cr  &\qquad
L_{k} = \sum_{r \in {\bZ} + {\half}}  r\  : {\tilde\psi}_{r}\psi_{r+k}  :
\cr}}
The importance of the fermions is the relation between the partitions
and the excited states in the fermion Fock space:
\eqn\chst{\eqalign{&
\vert M ; {\l} \rangle = {\psi}_{-M + \frac{1}{2} -
{\l}_{1}} {\psi}_{-M +\frac{3}{2} - {\l}_{2}} {\psi}_{-M + \frac{5}{2} - {\l}_{3}}
\ldots {\psi}_{-M - \frac{1}{2} + i - {\l}_{i}} \ldots
\cr }}
The operators ${\bW}_{k}$ introduced in \wgen\ are diagonal in the basis
of "partition" states:
\eqn\wgeneig{
{\bW}_{k} | M ; {\l} \rangle =
{1\over \hbar k}{\rm ch}_{k}( {\hbar}M, {\l} ) | M ; {\l} \rangle
}
with the eigenvalues given by \chnrp. In what follows, we shall also use
the formula \uchebnik\ref\op{A.~Okounkov, R.~Pandharipande, {\it The equivariant
Gromov-Witten theory of ${\bf P}^1$}, math.AG/0207233; {\it Gromov-Witten theory,
Hurwitz theory, and completed cycles}, math.AG/0204305}:
\eqn\nashesos{e^{J_{-1}\over {\hbar}} \vert M ; {\emptyset} \rangle
= \sum_{\l} {\bm}_{\l} {\hbar}^{-\vert {\l} \vert} \vert M ; {\l} \rangle}
and its direct consequence
\eqn\nashevashe{ \tilde{\psi}_{-r}\ e^{J_{-1}\over \hbar} \vert M+1 ; {\emptyset}
\rangle
= \hbar^{-r-\half}\sum_{{\l}} {{\bm}_{{\l}}\over {\hbar}^{|{\l}|}}\
\prod_{i=1}^{\infty}
{ i - {\l}_{i} + r-\half-M \over i } \ \vert M ; {\l} \rangle
}
where the sum is now taken over all partitions do not containing
an eigenvalue, corrseponding to $r$-th fermionic mode, which is automatically
taken into account vanishing of the product factor.
The infinite product in \nashevashe\ is actually finite
\eqn\infprd{
\prod_{i=1}^{\infty}
{ i - {\l}_{i} + r-\half-M\over i }   = {1\over \Gamma(r+\half-M)}
\prod_{i=1}^{{\ell}_{\l}}
{ i - {\l}_{i} + r-\half-M\over i +r-\half-M}
}

\vskip 15pt
\ndt
{\it Baker-Akhiezer functions}

\ndt
The sum over partitions \zuone\ can be compactly written as a matrix element in the
theory of free fermions \lmn\nikand:
\eqn\zfun{
Z ( a , {\bt}, {\hbar} ) =
\langle M \vert e^{-{J_{1}\over \hbar}} e^{{1\over\hbar}\sum_{k>0}
t_{k} {\bW}_{k+1}} e^{J_{-1} \over \hbar} \vert M \rangle =
\sum_{{\l}}
{{\bm}_{\l}^2\over{( - {\hbar}^2)^{| {\l} |}}} \
e^{ {1\over {\hbar}^{2}} \sum_{k>0}
t_{k}{{\ch}_{k+1} ( a, {\l})\over k+1}}
}
where $a = {\hbar}M$ for $M\in {\bZ}$, and the second equality follows from
\nashesos, \wgeneig.
Let $r \in {\bZ} + {\half}$ and set $z = {\hbar}r$.
Consider the Baker-Akhiezer function, or the following matrix element:
\eqn\mmelm{
{\tilde\Psi} ( z, a, {\bt}, {\hbar}) = {1\over Z ( a , {\bt}, {\hbar} )}
\langle M \vert  e^{-{J_{1}\over \hbar}} \ {\tilde\psi}_{-r} \
e^{{1\over\hbar}\sum_{k>0} t_{k} {\bW}_{k+1}}\
e^{J_{-1} \over \hbar} \vert M+1 \rangle
}
Using \nashesos, \nashevashe\
it can be expanded into a sum over partitions
\eqn\psipll{\eqalign{
{\tilde\Psi} ( z , a , {\bt}, {\hbar} )
\ =\  & \
%( - {\hbar})^{1- p}
{\hbar^{-r-\half}\over Z ( a , {\bt}, {\hbar} )}
\ \exp{1\over\hbar^2}\sum_{k>0} {t_{k}\hbar^k\over k+1}
\left((r+\half)^{k+1}-(r-\half)^{k+1}\right)\cdot \cr
 &  \cdot \ \sum_{\l} {\bm
}_{\l}^{2} ( - {\hbar}^{-2} )^{\vert {\l} \vert}
e^{{1\over\hbar^2}\sum_{k>0} t_{k} {{\rm ch}_{k+1}(a,{\l})\over k+1}  }
\prod_{i=1}^{\infty}
{ i - {\l}_{i} + r-\half-M\over i }
\cr}}
The latter can be obtained from the partition
function \zfun\ by the shift of times $t_{k} \to t_{k} - {\d}_{k}(\hbar,r)$:
\eqn\zfunsht{\eqalign{ &
{\tilde\Psi} ( z , a , {\bt}, {\hbar} )
\ =\  \ {\hbar}^{-r-\half}  \
\exp{1\over\hbar^2}\sum_{k>0} {t_{k}\hbar^k\over k+1}
\left((r+\half)^{k+1}-(r-\half)^{k+1}\right)\ \cdot \cr
& \qquad\qquad\qquad\qquad\qquad\qquad\qquad\qquad
\cdot\ {e^{M\Gamma'(r+\half)/\Gamma(r+\half)}\over\Gamma(r+\half)}\
{Z ( a, {\hbar}, {\bt} - {\bf \d}) \over
Z ( a, {\hbar}, {\bt})}
\cr }}
%where the coefficients $\delta = \{ \delta_k(\hbar,r)\}$ can be
defined via the generation function
\eqn\dellog{
{1\over \hbar^2}\sum_{k>0}  {\d}_{k} {x^{k+1}\over k+1}
= {\rm log}  {{\Gamma} \left(  r +\half- { x\over\hbar}\right) \over
{\Gamma} \left(  r + \half \right) }
+ {x\over\hbar}\log{\Gamma'(r+\half)\over\Gamma(r+\half)}
}
{}From \zfunsht\ one gets, that asymptotically at $\hbar\to 0$ and $\hbar r=z\to\infty$
\eqn\quasf{
{\tilde\Psi}(z, a, {\bt}, {\hbar}) \sim \exp{ S(z, a, {\bt}) \over \hbar}
 }
with the singularity
\eqn\quasymp{
S(z, a, {\bt}) = \sum_{k>0}t_kz^k - z(\log z-1) + a\log z + \ldots
}
while the nonsingular at $z\to\infty$ terms in \quasymp\ are expressed, as follows from
\zfunsht, in terms of the first  derivatives of $\log\ Z ( a , {\bt}, {\hbar} )$
w.r.t. ${\bf t}$-variables. The essential singularity \quasymp\ contains the
Eguchi-Yang term $z ({\log}z - 1)$, (cf. \egu), coming from the
Gamma function $\log \Gamma (z) \sim z(\log z-1)$ in \zfunsht. The fact that
the regular term of the first order in $z$, $t_1 z$ in \quasymp\ is accompanied
by the $ z ( {\log} z - 1)$ term, of the degree $1 + {\e}$, in a sense, is the
consequence of the asymptotic freedom of the noncommutative $U(1)$ theory
(see \nikand\ for the discussion of this phenomenon). In the $U(N)$ case we
shall see the analogous shift,
$t_1 \to t_1 - N {\log}z$, in agreement with the exact coefficient, $N$,
of the beta function of $U(N)$ super-Yang-Mills theory
\ref\nsvz{V.A.~Novikov, M.A.~Shifman, A.I.~Vainshtein, V.I.~Zakharov,
{\it "Beta function in supersymmetric gauge theories: instantons versus
traditional approach"}, \plb{166}{1986}{329-333}, Sov.J.Nucl.Phys. {\bf 43} (1986) 294,
{\cyr Yad.Fiz.} {\bf 43} (1986) 459-464}.

One can also study the fermion matrix element in the coordinate
representation:
\eqn\fermc{\eqalign{&
{\hat\Psi}(w, a, {\bt} , {\hbar} ) = {1\over Z ( a , {\bt}, {\hbar} )}
\langle M \vert  e^{-{J_{1}\over \hbar}} \
e^{{1\over\hbar}\sum_{k>0} t_{k} {\bW}_{k+1}}\  {\tilde\psi} (w) \
e^{J_{1} \over \hbar} \vert M +1\rangle = \cr
& \qquad\qquad\qquad\qquad\qquad = \sum_{r \in {\bZ} + {\half}} w^{-r}
{\tilde\Psi} \left(r\hbar, a, {\bt}, {\hbar}\right)
\cr }}
Asymptotically, \fermc\ gives
\eqn\quasff{
{\hat\Psi} (w, a, {\bt}, {\hbar})
 \sim \exp{{\Sigma} (w, a, {\bt}) \over {\hbar}}
 }
with the quasiclassical phase:
\eqn\quasph{
{\Sigma} (w, a, {\bt})  = S(z, a, {\bt}) - z \ {\rm log} w, \qquad
%{\Phi}(z) \equiv
{dS(z, a, {\bt})\over dz} = {\log} w}
related to \quasf\ by the Legendre transform, while \fermc\ can be thought as
an integral duality transformation \pqdual\ for the Baker-Akhiezer functions.
 Note (for fixed $(a, {\bt})$):
\eqn\swdf{d{\Sigma} = - z {dw \over w}}

The nonabelian theory partition function also has a representation in terms
of free fermions. For the gauge group $U(N)$ one uses $N$ flavours,
${\psi}^{(l)}(w)$, $l = 1, \ldots , N$. For the special values of the vacuum
expectation value of the Higgs scalar ${\Phi}$, $a_{l} = {\hbar} \left( M_{l}
+ {\rho}_{l} \right)$, where $M_{l} \in {\bZ}$ are integers, ${\rho}_{l}= {1\over N}\left( N+1 - 2 l \right)$,
the partition
function of the nonabelian theory
can be related to that of the abelian one by means of the procedure of
{\it blending} of partitions. In the language of free fermions this is the
following procedure:
\eqn\blefr{{\Psi} ( w^{1\over N} )  = {1\over\sqrt{N}}
 \sum_{l=1}^{N}  {\psi}^{(l)} (w) w^{-{\rho}_{l}}}
The single free fermion ${\Psi}$ is produced out of $N$ free fermions
${\psi}^{(l)}$, $l = 1, \ldots, N$ living on the $w$-space. The fermion
$\Psi$ lives on the $N$-fold cover of the original space, which
in quasiclassical limit turns into a spectral curve.

The spectral curve for a nonabelian theory will be explicitly described below.
On a small phase space (with only $t_1$ nonvanishing) it coincides with the
Seiberg-Witten curve, but when higher Casimirs are switched on, it goes beyond
the Seiberg-Witten family, still being a hyperelliptic curve for the $U(N)$
gauge theory. This curve corresponds to a quasiclassical hierarchy \kriw\ of
the Toda type, and we shall see that co-ordinate $z=\hbar r$, corresponding to
the Baker-Akhiezer functions of the type \quasf\ is distinguished from the point
of view of this hierarchy.

\subsec{Genus expansion}

The partition function \zuone, \zunntre\ has the same genus expansion, as
in \qcorr, but with the switched on times ${\bf t}$.
%\eqn\genexp{
%Z( a, {\hbar} , {\bt} ) = {\exp} \sum_{g=0}^{\infty}
%{\hbar}^{2g - 2} F_{g}( a, {\bt} )}
There are various interpretations of this expansion, which will be presented
in detail elsewhere \niklec.

\item{{\cyr a}$.\rangle$} Gauge theory interpretation in four dimensions:
the $\Omega$-background acts as a "smart box" with parameter ${\hbar}^{2}$
being the inverse volume, prepotential
as the extensive part of the free energy,
the higher terms $F_{g}$ being the finite size effects.

\item{{\cyr b}$.\rangle$} Topological string interpretation: in the geometrical
engineering setup
 $F_{g}$'s are the genus $g$ topological string amplitudes on a local Calabi-Yau
 manifold.

 \item{{\cyr v}$.\rangle$} Van-der-Vaals gas in one dimensions: the
 $\hbar$-expansion becomes the Mayer diagram expansion \niklec\ in this case.

\item{{\cyr g}$.\rangle$} We can view
\csacs\ as the quantum field theory  and perform the standard Feynman
diagram expansion around the Gaussian action $ {1\over 4{\pi}} \int_{{\bS}^{2}}
{\p\varphi}{\pb\varphi}$ with the vertices of valency $p$ weighted
with the weight ${\hbar}^{p-2}$, so that the sources
$\oint {\p\varphi} w^{\pm 1}$ have the weight ${\hbar}^{-1}$,
the "mass term" $t_{1}\oint  \left( {\p\varphi} \right)^{2}$ having no
$\hbar$-dependent factor at all, etc. The weight of the Feynman diagram
${\gamma}$ with $v_{p}({\g})$ vertices of valency $p$ would be, then
$$
{\hbar}^{\sum_{p} ( p - 2) v_{p}({\g}) } = {\hbar}^{2 ( E({\g}) - V({\g})) }=
{\hbar}^{2g-2}
$$
where
$$
E({\g}) = {\half} \sum_{p} p v_{p}({\g})
$$
is the number of edges, and $$V({\g}) = \sum_{p} v_{p}({\g})$$
is the number of vertices, and $g$ is the number of loops.
In this way one can make contact with the string loop expansion.

\subsec{The quasiclassical solution}
The partition functions \zuone\zunntre\ should be viewed as the partition
functions of an instanton gas in the "box" of size ${\hbar}^{-2}$.
We are interested in the free energy of the instanton gas, per unit volume:
\eqn\freeen{{\CF} ( {\vec a}, {\bt} ) = {\rm Lim}_{{\hbar} \to 0} \
{\hbar}^{2} {\rm log}Z  ( {\vec a} , {\hbar} , {\bt}) }
\vskip 15pt
\ndt
{\it Energy and entropy}

\ndt
The evaluation of the asymptotics \freeen\ is greatly facilitated by the
observation that in the limit where $\hbar \to 0$ with the rest of the parameters
kept fixed, the sum over the partitions $\l$ (or $N$-tuples $\vec \l$) is
dominated by the single partition of the very large size $\sim \hbar^{-2}$ \nikand.
To determine this partition and to evaluate \freeen\ one can use the geometric
representation of partitions, Young diagrams. In the limit $\hbar \to 0$ the
boundary of Young diagram of the master partition becomes a continuous curve.
This curve, e.g. the profile $f_{\l}(x)$, can be found by maximizing the weight
of the partition $\l$ in the sum $\zuone$.
The Plancherel measure \planchf\ and the Chern characters
\chnrp\ in \zuone\ combine together into the following {\it energy functional}
\eqn\energ{E ( {\l}, {\hbar} ) = \frac{1}{4}\int_{x_{1}>x_{2}} dx_{1}dx_{2} \ f''_{\l}(x_{1})\
f''_{\l}(x_{2}) \ {\gamma}_{\hbar} (x_{1}-x_{2}) -
\frac{1}{2\hbar^2} \int dx \ f''_{\l}(x) {\bt}(x)  }
so that
\eqn\zuinen{Z ( a, {\bt}, {\hbar}) = \sum_{\l} e^{-E ( {\l}, \hbar )}}
Expanding the energy $E({\l}, \hbar)$ in $\hbar$ (assuming $f_{\l}(x) \to f(x)$):
$$
E({\l}, \hbar) \sim {1\over {\hbar}^2} {\CE}[f]
$$
where
\eqn\enrgen{{\CE} [ f] = \frac{1}{4}\int_{x_{1}>x_{2}} dx_1 dx_2 \ f''(x_{1})\ f''(x_{2})
\ {\bF} (x_{1}-x_{2})
- \frac{1}{2} \int dx \ f''(x) {\bt}(x) \ ,}
we can approximate the partition function as:
\eqn\zuineni{Z ( a,  {\bt}, {\hbar} ) \sim {\exp} \left( - {1\over\hbar^{2}}{\CE}[f] \right) }
Indeed, the entropy contribution, i.e. the number of partitions $\l$, whose
profiles $f_{\l}(x)$ differ from $f(x)$ by $O ( {\hbar} )$, is of the order
of
$$
{\exp} \left({1\over\hbar} L\right) \ ,
$$
where $L$ is finite in the $\hbar \to 0$ limit. Thus in the $\hbar \to 0$ limit the entropy contribution is negligible, compared to that of the energy.

\vfill\eject
\ndt
{\it The variational problem}

\ndt
Thus, we arrive at the following conclusion:
\vskip 15pt
\ndt The quasiclassical tau-function or generalised Seiberg-Witten
prepotential is the critical value of the functional
\eqn\functnl{
\mathboxit{{\CF}( {\vec a}, {\bt})  = -\ {\rm Crit}_{f}\
{\CE} [ f ]}\ ,}
where we look for the extremum of the energy functional ${\CE}[f]$
in the class $C^{1}({\bR})$ of differentiable functions $f(x)$, such that
\eqn\ffcndt{\mathboxit{\eqalign{& -N \leq f'(x) \leq N \cr
& \qquad f''(x) \geq 0 \cr}}}
For $\vec a$ in appropriate domain of the moduli space $\CM$ the support of $f''(x)$ is a set of $n$ disjoint intervals
$\{ {\bI}_{l} \}_{l = 1, \ldots , n}$ along the real axis:
\eqn\suppsplt{\mathboxit{\eqalign{&
\qquad\qquad\quad {\rm supp} f'' = \amalg_{l=1}^{n} \ {\bI}_{l}, \cr
& \qquad\qquad\qquad {\bI}_{l} = \left(  x_{l}^{-} , x_{l}^{+} \right)  \  , \cr
& x_{1}^{-} < \ldots < x_{l}^{+} < x_{l+1}^{-} < x_{l+1}^{+} < \ldots x_{N}^{+} \cr}}}
The moduli ${\vec a}$ enter the variational problem via the additional constraints:
\eqn\fxda{\mathboxit{
a_{l}
= \frac{1}{2} \int_{{\bI}_{l}} dx\ xf''(x) \ ,}}
where ${\bI}_{l}, \ l = 1, \ldots , n$ is the $l$'th connected component of
${\rm supp} f''$.

The number $n$ of cuts, and independent moduli $a_l$ depend
on the gauge group. For example, for $G = U(N), SU(N)$, $n = N$.
For $G = SO(N)$, $n = 2\left[{N \over 2} \right]$, and for $G = USp(2N)$
there are $n = 2N+1$ cuts.

In the $G = SU(N)$ case the Higgs eigenvalues $a_l$ obey:
$\sum_{l=1}^{N} a_l = 0$. In the $G=SO(N)$ case the Higgs eigenvalues
have the form: $\left( a_{1}, a_{2} , \ldots, a_{n/2}, - a_{1}, - a_{2}, \ldots ,
-a_{n/2} , * \right)$, where for odd $N$, $* = 0$ and for
$N$ even $*$ is absent. In the $G= USp(2N)$ case the Higgs eigenvalues
 have the form: $\left( a_{1}, a_{2} , \ldots, a_{N}, - a_{1}, - a_{2}, \ldots ,
-a_{N} , 0 , 0\right)$.

It is easy to notice \amv\ that our variational problem is very similar to arising
in the context of multisupport solutions for matrix integrals.
However,
there are important distinctions, caused basically by properties of the functions
\ffcndt, which are extremising the functional \functnl. Below we
present the solution to the variational problem in the same geometric terms, though
involving
sometimes the multivalued differentials on hyperelliptic curves. We shall return to
the parallels between our approach and the matrix models, when discussing the
Eguchi-Yang matrix integral.

\newsec{Solution in the $U(1)$ case}

In this chapter we study in  detail the extended prepotential in the case of
$U(1)$ theory.
In the $U(1)$ case,  the constraint
\fxda\
can be in standard way taken into account with the help of the
Lagrange multiplier
\eqn\dfunctnl{
{\CE} [f]  \rightarrow {\CE}[f , a] = {\CE}[f] - a^{D}\left(
a-\frac{1}{2}\int dx\ x\ f^{\prime\prime}(x)\right)}
The Lagrange multiplier $a^{D}$ naively looks like the zeroth time $t_{0}$, since
its contribution can be viewed as the shift ${\bt} (x) \to {\bt} (x) + a^{D} x$.
However, in our setup (as often happens for similar problems, see e.g.
\ref\zabr{P.~B.~Wiegmann, A.~Zabrodin,
{\it ``Conformal maps and dispersionless integrable hierarchies,''} hep-th/9909147,
  Commun.\ Math.\ Phys.\  {\bf 213} (2000) 523\semi
   I.~K.~Kostov, I.~Krichever, M.~Mineev-Weinstein, P.~B.~Wiegmann, A.~Zabrodin,
  {\it ``$\tau$-function for analytic curves,''} hep-th/0005259}
  \ref\diri{A.~Marshakov, P.~Wiegmann, A.~Zabrodin,
  {\it ``Integrable structure of the Dirichlet boundary problem in two
  dimensions,''} hep-th/0109048, Commun.\ Math.\ Phys.\  {\bf 227} (2002) 131\semi
  I.~Krichever, A.~Marshakov, A.~Zabrodin,
  {\it ``Integrable structure of the Dirichlet boundary problem in
  multiply-connected domains,''} hep-th/0309010,
  Commun.\ Math.\ Phys.\  {\bf 259} (2005) 1}) the variable $a$ is fixed,
while $a^{D}$ is varied, thus
the dual variable $a$ plays the role of the zeroth time.
For $N > 1$ the situation is more complicated and will be discussed later on.

\subsec{The limit shape}

The variational equation for the functional
\dfunctnl\  gives
\eqn\exeq{
- {\half} \int d{\tilde x}\ f^{\prime\prime}({\tilde x})\ (x-{\tilde x})\left(\log |x-{\tilde x}|-1\right)
+ {\bt}'(x)  = a^{D} \ \qquad x \in {\rm supp} f^{\prime\prime}
}It is convenient to introduce the following analytic multivalued function
\eqn\Sfun{
S(z) = - \half \int dx \ f^{\prime\prime}(x)\ (z-x)\left(\log(z-x)-1\right)+
{\bt}^{\prime} ( z) }
on the $z$-plane with the ${\rm supp}f^{\prime\prime}$ removed. It
has the following properties:

\item{1.$\rangle$}
The differential ${\Phi} ( z ) =dS/dz$ is multivalued. However, the
  differential $d\Phi$ is already well-defined on $\CC$, the double cover
  of the $z$-plane, which is ramified at the end-points of the interval
  of the support of $f^{\prime\prime}$.

\item{2.$\rangle$}
The exponential $\exp\left({\Re \rm e}\ \Phi\right)$ is therefore single-valued
  on $\CC$.  On the cut it is equal to one.

\item{3.$\rangle$}
The equation \exeq\ implies that  the real part
$$\Re {\rm e} S(x) = \half\left(S(x+i0)+S(x-i0)\right) = a^D$$
is constant on the cut ${\bf I}={\rm supp}f^{\prime\prime}$.
In order to consider the asymptotic of \Sfun\ in what follows we shall always
  choose a branch, which is real along the real axis.

\item{4.$\rangle$}
Asymptotically, as $z \to \infty$,
\eqn\sasympt{
S(z)\ \ {=}\ \ - z\ \left({\log} z-1\right) + a\ {\log} z
+ \sum_{k=1}^{\infty} t_{k} z^{k}  - \sum_{k=1}^\infty {1\over k z^k}{{\p}{\CF}
\over {\p} t_k}}
where, according to \functnl,
\eqn\derF{
{\p {\CF}\over {\p} t_k} = {1\over 2(k+1)}\int dx\  x^{k+1}  f''(x)\ ,\ \ \ \ k>0}
and the coefficient in front of the $z(\log z -1)$ term is fixed by
\eqn\normlog{
\int dx \ f''(x) = f'( + \infty)-f'( -\infty)=2}

\item{5.$\rangle$} The Legendre transform of $S(z)$, ${\Sigma}(w)$,
\eqn\legds{d{\Sigma} = - z d{\Phi} =
d \left( {\bt}'  - z {\bt}'' \right) - {z dz \over 2} \int {dx \ f''(x) \over z - x}  }
expands near $z = \infty, w = \infty$ as follows:
\eqn\legdsexp{{\Sigma}(z)  = -z - \sum_{k=2}^{\infty} (k - 1 ) t_{k}z^{k}
- a \ {\log}z  + \sum_{k=1}^{\infty} { k +1 \over k z^{k}} {{\p\CF}\over {\p t_{k}}}}
The formulae \derF\ together with the equation
\eqn\aad{
a^D = {\p {\CF}\over \p a} \ , }
which follows from \functnl,
identify the generating function \functnl\ with the logarithm of quasiclassical
tau-function, being in the case of the single cut, a tau-function of dispersionless Toda
chain hierarchy. Note, that
the asymptotics \sasympt\ follows from \psipll\zfunsht.
Note also that the term, linear in $z$ in \legdsexp, can be viewed as the
regularised  $k=1$ term $(k-1)t_{k}z^{k}$,
where $t_{1}$ is replaced by
the divergent term ${\tau}_{0} + t_{1}$. The bare coupling ${\tau}_{0}$
is a logarithmically divergent function of the energy cutoff \nikand.

\item{6.$\rangle$} The pair of multi-valued functions $({\Phi}(z,a), z)$
can be viewed as the quasiclassical analogues of the pair of Orlov-Shulman and
Lax operators \taktak. The canonical transformation,
taking the pair $({\log} w , a)$ to $( {\Phi},  z)$ is generated by the
generating function $S(z, a)$, or, equivalently, ${\Sigma}({\Phi},a)$.

\subsec{Toda as presentiment}

As we have already discussed, the limit shape is described by the
function $f(x)$, which differs from $| x - a |$ on the single interval
${\bI}  = ( x^{-}, x^{+})$.
The corresponding integral transforms $S(z)$ and
${\Phi}(z)$ suggest to study the double cover ${\CC}$ of the
$z$-plane with the branch points $z=x^{+}$ and $z=x^{-}$.

The curve ${\CC}$ can be equivalently described by the equation
\eqn\uIcurve{
z=v+\Lambda\left(w+{1\over w}\right)}
in ${\bC} \times {\bC}^{*}$,
with $x^{\pm} = v\pm 2\Lambda$. It is just a copy of a Riemann sphere or ${\bC\bP}^1$,
with two marked points $P_\pm$, with
$z(P_\pm)=\infty$, $w^{\pm 1}(P_\pm)=\infty$.
%Sometimes we use the notation ${\infty}_{\pm} = P_{\pm}$.
Another form of the curve \uIcurve\ will be generalised
in the nonabelian case:
\eqn\uIcurveii{y^2 = ( z - v)^2 - 4 \Lambda^2 = ( z - x^{+} ) ( z - x^{-})}
where $y = {\Lambda} \left( w - w^{-1} \right)$. The formula \Sfun\ defines a function
with a logarithmic cut and asymptotic behaviour \sasympt, odd under the
involution $w\leftrightarrow {1\over w}$ of the curve \uIcurve. In terms
of the variable $w$ one can globally write
\eqn\Sw{
S = - S_{EY}(w) +
\sum_{k>0} t_k\Omega_k(w) + a\ \log w + a^D \ ,}
with the Eguchi-Yang term
\eqn\ey{
S_{EY}(w) = \left(v+\Lambda\left(w+{1\over w}\right)\right)\log w
+ \Lambda(\log\Lambda-1)\left(w-{1\over w}\right) \ ,}
the "Hamiltonians"
\eqn\Ow{
\Omega_k (w) = z^k_{+}-z^k_{-},  \ \ \ k>0}
which are the Laurent polynomials in $w$, and the "zeroth Hamiltonian"
\eqn\OO{
\Omega_0 (w) = \log w \ , }
all odd under $w\leftrightarrow{1\over w}$.
The Eguchi-Yang term \ey\ is fixed by its asymptotics
$S_{EY}(w) = \pm z(\log z-1) + O(1)$ at $P_\pm$ (cf. with \quasymp).
Note that one can write the differentials $d{\Omega}_{k}$ in the following
suggestive form:
\eqn\Owii{
d{\Omega}_{k} = k  \left( z^{k-1} y \right)_{+} {dz \over y}}
where the symbol $(\ldots)_{+}$ denotes now a polynomial part in $z$, in the
expansion near $P_{+}$.
We can think of \Sw\ as of the Legendre transform of the Seiberg-Witten differential
$d\Sigma = -z{dw\over w}$. Obviously, on the cut \Sw\ gives $\left.S\right|_{w=1} =
a^D$.
{\sl The canonical Toda chain times are defined by the residues
\eqn\tOres{
t_0 = {\rm res}_{P_+} dS = - {\rm res}_{P_-} dS = a}
and
\eqn\tP{
t_k = {1\over k}\ {\rm res}_{P_+} z^{-k}dS =
- {1\over k}\ {\rm res}_{P_-} z^{-k}dS,\ \ \ k>0}}
{}From the expansion \sasympt\ at $z\to\infty$
it also immediately follows, that
\eqn\tPd{
{\p {\CF}\over \p t_k}
%= {1\over k+1}\int dx f''(x) x^{k+1}
= {\rm res}_{P_+} z^{k}dS
= - {\rm res}_{P_-} z^{k}dS,\ \ \ k>0}
Notice that \sasympt\ has no constant in $z$ term, which together with
explicit formula \Sw\ allows to compute $a^D$ from a regularized value
of the function \Sw\ at $z\to\infty$
\eqn\aD{\eqalign{ &
a^D = \left.S\right|_{w=1} = \cr
& = \lim_{z\to\infty}\left(\sum_{k>0}t_{k} z^{k} - z(\log z-1) + a\log z
+S_{EY}(w) -
\sum_{k>0} t_k\Omega_k(w) - a\log w\right)
\cr } }
The consistency condition for
\tPd\ is ensured by the symmetricity of the second derivatives
\eqn\sysi{
{\p^2 {\CF}\over \p t_n\p t_k} = {\rm res}_{P_+} (z^k d\Omega_n)}
where we have introduced
\eqn\Oz{
 {\p S\over\p t_k}= \Omega_k \ {=}\ \pm\left(z^k
- {\p^2  {\CF}\over\p a\p t_k} -
\sum_{n>0}
{\p^2 {\CF}\over\p t_k\p t_n}{1\over nz^n}\right),\ \ \ k>0
}
$$
{\p S\over\p a}\ {=}\ \Omega_0 = \pm \left( \log z
- {\p^2 {\CF} \over\p a^2} - \sum_{n>0}
{\p^2 {\CF} \over\p a\p t_n}{1\over nz^n}\right)
$$
The functions $\Omega_k$, $k > 0$, form a basis in the space of meromorphic
functions with poles at the points with
$P_{\pm}$, odd under the $w \mapsto w^{-1}$ involution. All time-derivatives
here are taken at constant $z$.

The expansion \Oz\ of the Hamiltonian functions \Ow\ expresses the second
derivatives of $\CF$ in terms of the coefficients of the curve \uIcurve, e.g.
\eqn\Omexp{\eqalign{ &
\Omega_0\ {=}\ \log z - \log\Lambda - {v\over z}
- {\Lambda^2+{v^2\over 2}\over z^2} + \ldots \cr
& \Omega_1\ {=}\ z - v - {2\Lambda^2\over z} - {2v\Lambda^2\over z^2}
+ \ldots \cr
& \Omega_2\ {=}\ z^2 - (v^2+2\Lambda^2) - {4v\Lambda^2\over z} -
{2\Lambda^2(\Lambda^2+2v^2)\over z^2} + \ldots \cr }}
as $z\to\infty$,
which gives, in particular,
\eqn\logt{
{\p^2 \CF \over \p a^2} = \log\Lambda,\ \ \
{\p^2 \CF \over \p a\p t_1} = v}
and
\eqn\todaeq{
{\p^2 \CF \over \p t_1^2} =2\Lambda^2=2\exp \left(2{\p^2 \CF \over \p a^2}\right)}
which becomes the long-wave limit of the Toda chain equations
after a derivative with respect to $a$ is taken:
\eqn\todaequ{{{{\p}^2 a^D}\over {{\p t_{1}^2}}} \ = \ 2 {\p \over {\p a}}
\exp \left(2 {\p a^D \over \p a}\right)}
with the Toda co-ordinate $a^{D} = {{\p\CF}\over{\p a}}$. The other expansion coefficients \Omexp, with the help of \sysi, give rise
to the Losev-Shatashvili polynomials.  In the $U(1)$ case they are  polynomials in the single variable
$u=v$, whose coefficients depend on $\Lambda$.

One can now find the dependence of the coefficients of the curve \uIcurve\
on the deformation parameters ${\bf t}$ of the microscopic theory by
computing $dS$ at the ramification points $w = \pm 1$, where
$z = v \pm 2\Lambda$ and $dz=0$:
\eqn\eqscu{
\left.{dS\over d\log w}\right|_{w =\pm 1} =
\sum_{k>0}t_k \left.{d\Omega_k\over d\log w}\right|_{w =\pm 1}
+ a-v \mp 2\Lambda\log\Lambda =0
}

\vfill\eject
\ndt
{\it Small phase space}

\ndt
If $t_k=0$ for $k>1$, solution to \eqscu\ immediately gives
\eqn\tOI{
v=a, \ \ \ \Lambda=e^{t_1}}
and the prepotential
\eqn\preptOI{
{\CF} = \ha {aa^D} +  \ha{\rm res}_{P_+}\left( zdS\right) - {1\over 4} a^2
= \ha\left(a^2t_1+e^{2t_1}\right)}
The slope $f^{\prime}(x)$ of the corresponding limit shape profile $f(x)$ is given by:
\eqn\fpp{f^{\prime}(x) = {2\over{\pi}} {\rm arcsin} \left( {x - a  \over 2e^{t_{1}}}
\right), \qquad x \in \left( a - 2 e^{t_{1}} , a + 2 e^{t_{1}} \right) }
where the branch of arcsin (hence the name of the "arcsin law" \kerovi)
is chosen so that $f^{\prime} \left( a \pm 2 e^{t_{1}} \right)  = \pm 1$.
The limit shape itself is given by the  function:
\eqn\fofx{f(x) = {2\over{\pi}} \left( \sqrt{4 e^{2t_{1}} - ( x - a)^2} + ( x- a)
{\rm arcsin}
\left( {x - a \over 2 e^{t_{1}}} \right) \right), \qquad x \in \left( a - 2 e^{t_{1}} ,
a + 2 e^{t_{1}}  \right)}
and
\eqn\fofxo{f (x ) = | x - a |, \qquad {\rm otherwise.}}

\vskip 15pt
\ndt
{\it Turning on $t_2$}

\ndt
Now let us turn on the Casimir ${\tr}\,{\Phi}^{3}$ in the ultraviolet, or
in other words consider nonvanishing $t_{1}, t_{2}$. Then one finds
from \eqscu
\eqn\tOII{ \eqalign{ &
v=a -{1\over 4t_2}{\bf L}\left(-16t_2^2e^{2(t_1+2t_2a)}\right) \cr
& \log\Lambda = t_1+2t_2a - \half{\bf L}\left(-16t_2^2e^{2(t_1+2t_2a)}\right)
\cr
}}
where ${\bL}(t)$ has an expansion:
\eqn\Lambd{
{\bL}(t) = \sum_{n=1}^{\infty} {(-n)^{n-1} \over n!} t^{n} =
t - t^2 + {3\over 2} t^{3} - {8 \over 3} t^{4} + \ldots}
and obeys the functional equation
\eqn\Lambert{
\bL (t)\  {\exp} \ {\bL} (t) = t}
For the prepotential one gets instead of \preptOI
\eqn\preptOII{ \eqalign{ &
{\CF} = \ha {aa^D} +  \ha{\rm res}_{P_+}\left( zdS\right)
-{t_2\over 2}{\rm res}_{P_+}\left( z^2 dS\right) - {1\over 4} a^2 = \cr
& = {\bf t}(a) + \ha e^{2{\bf t}''(a)} +
 2t_2^2\ e^{4{\bf t}''(a)} +  {64t_2^4\over 3}\ e^{6{\bf t}''(a)} +
{1024 t_2^6\over 3}\ e^{8{\bf t}''(a)} + \ldots
\cr
}}
where the instanton expansion is governed by the
parameter $q = e^{\tau_0}$ related with the bare coupling, which can be introduced by
$t_1\to t_1+\log\ q$. Here $t_k=0$ for $k>2$ and ${\bf t}(a)=t_1{a^2\over 2}
+t_2{a^3\over 3}$, so that ${\bf t}''(a)= t_{1}+2t_{2} a$.

{}The slope of the limit shape is again given by the trigonometric functions:
\eqn\lmshp{\eqalign{& f^{\prime}(x) =
{2\over {\pi}} \left( {\rm arcsin}\left( {x - v \over 2{\Lambda}} \right) + 2t_{2}
\sqrt{4 {\Lambda}^{2}-(x-v)^2} \right) \cr
& \qquad\qquad v - 2 \Lambda \leq x \leq v + 2 \Lambda}}
where $v$ and $\Lambda$ are the functions \tOII\ of $t_{1}, t_{2}, a$. Hence we see,
that the Vershik-Kerov ``arcsin" law is deformed, for nonvanishing $t_2$, by the
Wigner semicircle distribution.

{}Note that our "rules of the game" are such that the higher order times $t_{k}$,
$k > 1$ are considered to be the nilpotent parameters.
Had we viewed them as the ordinary perturbation parameters (and thus faced the possible
non-renormalisability concerns), the formula \lmshp\ would have suggested
a possibility for the interesting critical behaviour, similar to the one found in the two dimensional
Yang-Mills theory \ref\dk{M.~Douglas, V.~Kazakov, hep-th/9305047\semi
D.~Gross, A.~Matytsin, hep-th/9404004}: indeed, as soon as $t_{2}$ becomes sufficiently large, $|t_{2}| > t_{2}^{*}$,  so that
\eqn\critcnd{
4 t_{2}^{*}{\Lambda}(a, t_{1}, t_{2}^{*}) = 1}
the slope $f^{\prime}(x)$ reaches $+1$ before $x$ reaches the right
end of the cut $(v - 2\Lambda , v + 2\Lambda)$. The equation \critcnd\ is nothing but the convergence radius of the series \Lambd\ for the function ${\bf L}(t)$.  For $| t_2 | > t_{2}^{*}$ a new cut opens up
and the theory goes over to another phase, in particular the curve $\CC$ would grow in genus. Of course, the critical value
of $t_{2}$ is exponentially large $\sim {\Lambda}^{-1}$, like the Landau pole.

\subsec{Symplectomorphisms and dispersionless hierarchy}
The evolution of the
prepotential ${\CF} ( a, {\bt})$ with respect to the times $t_{k}$
is described by the Toda chain hierarchy. We start by reminding the Lax formulation of
this hierarchy \taktak. The flows can be described in the Hamiltonian fashion,
by interpreting the evolution to take place on the space of holomorphic functions
on ${\CP} = {\bC}^{*} \times {\bC}$.
We endow the space ${\CP}$ with the holomorphic symplectic form:
\eqn\holsum{{\varpi} = {dw \over w} \wedge da}
where $w \in {\bC}^{*}, a \in {\bC}$.
The $k$'th flow is governed by the Hamiltonian
\eqn\kham{{\Omega}_{k} (w, a) = z^{k}_{+} - z^{k}_{-}\ , \qquad k > 1}
where, as in \Ow\uIcurve\ the $\pm$ denote the part, holomorphic at
$w = 0$ or $w = \infty$, respectively, and $z$ is the Lax operator:
\eqn\lax{z ( w, a) = v ( a ) + {\Lambda} ( a ) \left( w + {1\over w} \right) \ , }
and
\eqn\zrham{{\Omega}_{0} = {\log}\, w}
The evolution is encoded in the equations:
\eqn\laxeq{\eqalign{\qquad\qquad {{\p}\over {\p}t_{k}} z ( w, a ) \ = \ &\biggl\lbrace{ z ( w, a) \ , \ {\Omega}_{k} ( w, a )  \biggr\rbrace} \qquad \Leftrightarrow \cr
{\p v(a) \over {\p}t_{k}}\  + & \left( w + {1\over w}\right)  {\p {\Lambda(a)}\over {\p}t_{k}}  \ = \ \cr
\ = \ & {\Lambda} ( a) \left( w - {1\over w} \right) {\p}_{a} {\Omega}_{k} - \left( v'(a) w + {\Lambda}'(a) \left( w^2 + 1\right) \right)  {\p}_{w} {\Omega}_{k} \cr }}Thus:
\eqn\frstflw{\eqalign{{\p v(a) \over {\p}t_{1}} = 2 \left( {\Lambda}^{2} (a) \right)^{\prime} \ , \ & {\p {\Lambda}(a) \over {\p}t_{1}} = {\Lambda} (a) v(a)^{\prime} \cr
{\p v(a) \over {\p}t_{2}} = 4 \left( v (a) {\Lambda}^{2}
(a) \right)^{\prime} \ , \ & {\p {\Lambda}(a) \over {\p}t_{2}} =
{\Lambda} (a) \left( v(a)^{2} + 2 {\Lambda}^{2}(a) \right)^{\prime} \cr}}
The first line in \frstflw\ implies the Toda equation \todaeq.
Let us discuss the geometry behind the equations \frstflw\laxeq. They define
a set of commuting symplectomorphisms of $\CP$. Infinitesimally these
symplectomorphisms look like:
\eqn\infsym{{\d}w = w {\p}_{a} {\Omega}_{k} {\d}t_{k} , \qquad
{\d} a = - w {\p}_{w} {\Omega}_{k} {\d} t_{k}}
Let us denote the finite time symplectomorphism by:
\eqn\symplft{g ( {\bt} ) : {\CP} \to {\CP} \ , \qquad
\left( W ( w, a; {\bt} ), \ A ( w, a; {\bt}) \right) = g({\bt}) ( w , a)}
so that $g ( 0) = {\rm id}$.
Now let us take the Lax operator, viewed as a function on $\CP$, and let it
evolve in the geometric way:
\eqn\laxev{\eqalign{ z (w, a ; {\bt}) \ = \ & g({\bt})^{*} z ( w, a; 0) =
z \left( W \left( w, a; {\bt} \right),
\ A \left( w, a; {\bt}\right) \right) =\cr
&  v \left( A \left( w, a; {\bt} \right) \right) +
{\Lambda} \left( A \left( w, a; {\bt} \right) \right) \left( W ( w, a; {\bt} ) +
{1\over W ( w, a; {\bt} )} \right)  = \cr
& v ( a ; {\bt} ) + {\Lambda} ( a ; {\bt} ) \left( w + {1\over w} \right)\cr}}
the last equality being  a consequence of the asymptotic behaviour of the
Hamiltonian vector fields \infsym\ as $w \to 0$ and $w \to \infty$.

We can also change the canonical variables, from $(w,a)$ to $(z, {\Phi})$, such that:
\eqn\wazphi{{d w \over w} \wedge da = dz \wedge d{\Phi}}
By substituting \lax\ into \wazphi\ we can express $\Phi$ via $w$ and $a$
(of course, \wazphi\ determines $\Phi$ up to an addition of a function of
$z$):
\eqn\phiwa{{\Phi}(w, a) = \int^{a} {da' \over \sqrt{ ( z(w,a) - v(a'))^2 - 4 {\Lambda}^{2}(a')}} , \qquad z (w, a)  =
v(a) + {\Lambda}(a) \left( w + {1\over w}\right)}
Finally, the symplectomorphism $g ( {\bt})$ can be described with the help
of a generating function (see \arnold\ for the definitions):
\eqn\genfns{\dd S = {\Phi} dz + {\rm log} w \ d a}
where we view $S$ as a function of $(z, a)$. We can think of $(z,{\Phi})$
as of the functions of $W, A$, i.e. they coordinatise the image of $g({\bt})$.
Thus $S$ is a function of $z, a, {\bt}$. We claim it coincides with \Sfun.

\newsec{The nonabelian case}

\subsec{The energy and surface tension}

In the nonabelian case, $G= U(N), N > 1$, similarly to \dfunctnl,
one has to add to the functional ${\CE}[f]$ several Lagrange multipliers, or
the so-called surface tension term:
\eqn\dule{\eqalign{& {\CE}[f, {\vec a}] =
{\CE}[f] - \sum_{l=1}^{N} a^{D}_{l} \left( a_{l}-{\half}\int_{{\bI}_{l}}
dx\ x f''(x)\right) = \cr
& ={\CE}[f] + \frac{1}{2} a^{D} \int x f''(x) dx +
\int \ dx\ {\s}( f^{\prime}) - \sum_{l=1}^{N} a_{l} a^{D}_{l} \cr } }
where
\eqn\suma{a^{D} = {1\over N} \sum_{l=1}^{N} a^{D}_{l}}
and ${\s}$ is a piece-wise linear function, which vanishes for $t$ outside the
interval $(-N, N)$, and equals ${\half} ( a^{D}_{l} - a^{D} ) ( t + N - 2l +2)
+ \sum_{m=1}^{l-1} ( a^{D}_{m} - a^{D} )$, for $t \in ( - N + 2l - 2, -N + 2l)$,
for $l = 1, \ldots, N$, so that:
\eqn\surftns{\frac{1}{2} a^{D} \int x f''(x) dx +
\int \ dx\ {\s}( f^{\prime})  =
{\half} \sum_{l=1}^{N} a^{D}_{l} \int_{{\bI}_{l}} \ x  \ f''(x) \ dx}
Define, as before:
\eqn\sfunn{S (z) = - {\half} \int \ dx \ f^{\prime\prime}(x) ( z - x)
({\log} (z - x) - 1) + {\bt}^{\prime}  ( z) }
and
\eqn\phifunn{{\Phi}(z) = {dS\over dz} = - {\half} \int \ dx \ f^{\prime\prime}(x)
{\rm log} ( z - x ) + {\bt}^{\prime\prime} (z )}

\subsec{The curve}

The variational equation, following from \dule, can be written as:
\eqn\vare{
S(x)=\half\left(S ( x + i0 ) + S ( x - i0)\right) = a_{l}^{D}\ ,
\ {\rm for} \ x \in {\bI}_{l}}
which implies:
\eqn\varphi{{\Phi} ( x + i0) + {\Phi} ( x - i 0) = 0 \ ,
\ {\rm for} \ x \in {\bI}_{l}}
Together with the obvious property
\eqn\ims{S(x \pm i 0 )  = a^D_{l} \mp {i{\pi} \over 2}
\int_{x}^{\infty} \ d{\tilde x} \ f''({\tilde x}) \ ( x - {\tilde x})\ ,
\ {\rm for} \ x \in {\bI}_{l} }
it implies that the differential
\eqn\dfi{d{\Phi} = S^{\prime\prime}(z) dz =
- {\ha} \left( \int \ dx \ {f''(x) \over z - x} \right) dz +
{\bt}^{\prime\prime\prime} dz}
is well-defined on the double cover ${\CC}$ of the $z$-plane, which is
ramified at the points $x_{l}^{\pm}$:
\eqn\dcN{y^{2} = \prod_{l=1}^{N} ( z - x_{l}^{+} ) ( z - x_{l}^{-} )
%= z^{2N} + \sum_{i=1}^{2N} p_{i} z^{2N-i}
}
Let us denote by $P_{\pm}$ the two preimages of the point $z = \infty$.
The differential $d\Phi$ is odd under
the involution $ y \mapsto - y$. As $z \to \infty$, on one of the sheets of
$\CC$, say near $P_{+}$,  the differential $d\Phi$ has the following asymptotics:
\eqn\dfas{d\Phi =
\sum_{k>1} k(k -1) t_{k} z^{k-2} dz -  N{dz \over z} -
a { dz\over z^{2}} -
\sum_{k>1} k  {{\p \CF} \over {\p t_{k-1}}}{ dz\over z^{k+1}}
}
These two properties allow to write:
\eqn\difff{d{\Phi} = \sum_{k\geq 0} s_{k} z^{k} {dz \over y} =
\sum_{k>1} k(k-1)t_k d\Omega_{k-1} - N d\Omega_0 -
2\pi i\sum_{j=1}^{N-1}d\omega_j}
where the differentials $d\Omega_k$ are
fixed by their asymptotics at $z\to\infty$
\eqn\Omas{
d\Omega_k\ {\sim}  \ \Biggl\lbrace\matrix{& \pm kz^{k-1}dz + O(z^{-2}),\ \ \ \ k>0 \cr & \cr
&
 \pm  {dz\over z} + O(z^{-2}),\ \ \ \ k=0}}
and vanishing $A$-periods
$$
\oint_{A_i}d\Omega_k = 0,\ \ \ k\geq 0,\ \ \ \forall\ i=1,\ldots,N-1
$$
while $d\omega_i$, $i=1,\ldots,N-1$ are
canonical holomorphic differentials normalized to the $A$-cycles, surrounding the
first $N-1$ cuts.

The coefficients $s_{k}$ and the ramifications points $x_{l}^{\pm}$
are to be determined from the conditions:
\eqn\diffre{\eqalign{& {1\over 2\pi i}
\oint_{P_{\pm}} z^{-k+1} d{\Phi} = \pm k ( k - 1 ) t_{k}, \qquad
k = 2, 3, \ldots \cr
& \ \ \ \ \ \ {\rm res}_{P_\pm} d\Phi = \mp N  \cr}}
directly following from \dfas,
\eqn\diffreq{\eqalign{&
- {1\over 2\pi i } \oint_{A_{l}} d{\Phi} =\ha \int_{{\bf I}_l} f''(x) dx =
\half\left(f'(x_l^+)-f'(x_l^-)\right)  = 1 \ , \cr
& \ \ \  \qquad {1\over 2\pi i } \oint_{B_{l}} d{\Phi} =  0,
\qquad l = 1, \ldots , N \cr}}
If $s(z)=\sum_{k\geq 0} s_{k} z^{k}$ is a polynomial of power
$N+K-2$ (in the case of nonvanishing times $t_1,\ldots,t_K$ up to the $K$-th order),
its higher $K$
coefficients are fixed by leading asymptotic
and the residue at infinity via \diffre, while
the rest $N-1$ coefficients can be determined from fixing the $A$-periods
by the upper line in \diffreq. Now it leaves only
$2N$ branch points $\{ x_j^\pm\}$ of the curve \dcN\ to be fixed from the
rest of \diffreq\ (vanishing of the $B$-periods),
and the Seiberg-Witten periods of the "dual differential"
\eqn\diffreqsw{
{1\over 2\pi i } \oint_{A_{l}} d{\Sigma}  = -
{1\over 2\pi i } \oint_{A_{l}} z d{\Phi}  = a_{l} , \qquad l = 1, \ldots , N }

Together with normalization of $\Phi(x^+_N)=0$ at the largest positive
ramification point, suggested by integral representation \phifunn,
 the second conditions of \diffreq\ can be equivalently written as
$2N$ {\it real} conditions
\eqn\psibranch{
\Re e \Phi (x_j^\pm)=0,\ \ \ \ j=1,\ldots,2N}
The condition \psibranch\ ensure that under variation at constant $z$
$$
\delta\left(dS\right) = \delta \left(\Phi dz\right) \simeq {\rm holomorphic}
$$
modulo the exact terms.

Since under rescaling:
\eqn\resc{\eqalign{ & z \mapsto {\m} z, \qquad y \mapsto {\m}^{N} y
\cr
& x_{l}^{\pm} \mapsto {\m} x_{l}^{\pm}, \qquad a_{l} \mapsto {\m} a_{l}, \qquad
 t_{k} \mapsto {\m}^{1-k} t_{k}
 \cr}}
the function ${\Phi}(z)$ acquires logarithmic corrections,
the $t_{1}$ dependence can be fixed by
\eqn\tonefi{
t_1 = {\rm res}_{P_+}\left(z^{-1}\Phi dz\right)
}
with the appropriately chosen branch of $\Phi$ at the point $P_+$.

The Lagrange multipliers
$$
a^D_l = {\partial{\cal F}\over \partial a_l},\ \ \ l=1,\ldots,N
$$
can be computed by a standard trick. From \vare,\ims\ it is clear that
\eqn\adper{
a^D_i-a^D_j = \int_{x_j^+}^{x_i^-} dS = {1\over 2}\oint_{B_{ij}} dS}
or
\eqn\gradF{
{\partial{\cal F}\over \partial a_i} = {1\over 2}\oint_{B_i} dS, \ \ \ i=1,\ldots,N-1}
and still
$$
a^D = {1\over N}\sum_{i=1}^N a^D_i =
{1\over N}\sum_{i=1}^N{\partial{\cal F}\over \partial a_i}
%= {1\over N}{\partial{\cal F}\over \partial t_0}
$$
For the other time-derivatives one can write
\eqn\Ftres{
{\p \CF \over \p t_k} = {\rm res}_{P_+} \left( z^k dS \right) =
- {1\over k+1}{\rm res}_{P_+} \left( z^{k+1}d\Phi \right)
}

\bigskip
\subsec{Small phase space and Seiberg-Witten curves}

 Now suppose:
  $${\bt}(x) = t_{1} {x^{2} \over 2} \ , \
$$
The conditions \dfas\diffreq\
 imply that $w = \exp \left(- {\Phi}\right)$ is a meromorphic function on
 $\CC$ with the only pole at $P_{+}$, where it behaves as:
 \eqn\wzasym{w \propto z^{N} + \ldots \ , \ }
 (recall that $z(P_{+})=\infty$) and which transforms  under the $y \mapsto - y$
 involution as $w \mapsto w^{-1}$,
 (for the switched on
higher times, such function would acquire an essential singularity at the points
where $z=\infty$, and the constraints \diffreq\ cannot be
resolved algebraically).

These conditions fix $w$ uniquely, up to a few parameters:
\eqn\swcurv{ {\Lambda}^{N} \left( w + {1\over w} \right) =
%z^{N} + u_{1} z^{N-1} + u_{2} z^{N-2} + \ldots + u_{N}
\prod_{l=1}^N (z - v_l)
\equiv P_{N} ( z) }
i.e.
$$
d{\Phi} = -P^{\prime}_{N}(z) {dz\over y}, \qquad y =  {\Lambda}^{N} \left( w - {1\over w} \right)
$$
The coefficients of $P_{N}(z)$ are determined by the periods \diffreq:
\eqn\swper{-{1\over 2\pi i}\oint_{A_{l}} z d{\Phi} = a_{l}, \ l = 1, \ldots , N}
Finally, the parameter ${\Lambda}$ is related to the time $t_{1}$ via:
\eqn\lam{{\Lambda}^{N} = e^{t_{1}}}
In this way one easily recovers the results of \nikand.

\subsec{Quasiclassical flows}

Let us now come back to the general formula \defprep\ and show how it can be derived
from our quasiclassical solution on large phase space, or for nonvanishing
microscopic times. Taking an extra derivative of \Ftres\ one gets formula
\sysi, which can be re-written on generic non-abelian curve \dcN\ as
\eqn\sysiW{
{\partial^2{\cal F}\over \partial t_n\partial t_m} =
{\rm res}_{P_+} (z^m d\Omega_n) =
{\rm res}_{{P_+}\otimes {P_+}}\left( z(P)^n z(P')^m W(P,P')\right)}
where we have introduced the bi-differential $W(P,P')=d_Pd_{P'}\log E(P,P')$,
$E(P,P')$ being the prime form \ref\fay{J.~Fay, {\it ``Theta-functions on
Riemann surfaces"},
Lect.Notes Math., Vol. {\bf 352}, Springer, New York, 1973},
with the only second order pole at diagonal and
vanishing $A$ periods. In
the inverse co-ordinates $z=z(P)$ and $z'=z(P')$ near the point $P^+$ with
$z(P^+)=\infty$ it has expansion
$$
W(z,z') = {dzdz'\over (z-z')^2}+\ldots =
\sum_{k>0} {dz\over z^{k+1}}\ d\Omega_k(z') + \ldots
$$
The bi-differential $W(P,P')$ can be related with the Szeg\"o kernel \fay
\eqn\fay{
S_e(P,P')S_{-e}(P,P') =  W(P,P') +
 d\omega_i(P)d\omega_j(P')
{\partial\over \partial \tau_{ij}}\log\ \vartheta_e( 0|\tau)}
which, for a half-integer characteristics $e \equiv -e$,
has an explicit expression on hyperelliptic curve \dcN\
\eqn\Sz{
S_e(z,z') = \frac{U_e(z) +
U_e(z')}{2\sqrt{U_e(z)U_e(z')}} \frac{\sqrt{dz dz'}}{z-z'} }
with
\eqn
\Ue{
U_e(z) = \prod_{j =1}^{N}\sqrt{ \frac{z -
x_{e^+_j}}{z - x_{e^-_j}}} }
Here $\{ x_{e^\pm_j} \}$ is a partition of the ramification points of \dcN\ into
two sets, corresponding to a given characteristic $e$. For example, on a small phase
space, when \dcN\ turns into the Seiberg-Witten curve \swcurv, there is a distinguished
partition $e=E$ with
\eqn\UE{
U_{E} (z) = \sqrt{P_{N}(z)-2\Lambda^N\over P_{N}(z)+2\Lambda^N}
}
Substitution of \fay\ into \sysiW\ gives
\eqn\resfay{\eqalign{ &
{\partial^2{\cal F}\over \partial t_n\partial t_m}
= {\rm res}_{{P_+}\otimes {P_+}}\left( z(P)^n z(P')^m S_e(P,P')^2\right) - \cr
& - {\rm res}_{P_+} \left(z^n d\omega_i\right)
{\rm res}_{P_+} \left(z^m d\omega_j\right){\partial\over \partial \tau_{ij}}
\log\ \vartheta_e (0|\tau) = \cr
&
= {\CC}_{nm}\left(x_{e^\pm_j}\right) -
{\partial^2 \CF \over\partial a_i\partial t_n}
{\partial^2 \CF\over\partial a_j\partial t_m}\
{\partial\over \partial \tau_{ij}}\log\ \vartheta_e( 0|\tau)
\cr }}
where for the Losev-Shatashvili polynomials one gets from \Sz
\eqn\poluC{
{\CC}_{nm}\left( x_{e^\pm_j} \right) = \ha {\rm res}_{{P_+}\otimes {P_+}}
\left({z^kz'^n\over (z-z')^2}\left(1+{U_e(z)\over 2U_e(z')}+
{U_e(z')\over 2U_e(z)}\right)dzdz'\right)
}
If calculated on the small phase space, where all $t_k=0$, with $k>1$,
and the particular choice of the characteristic \UE,
which correspond to the deformation \defprep\ around the Seiberg-Witten prepotential,
residues in \poluC\ vanish for $n,m<N$, and one gets
\eqn\rgsw{
{\partial^2{\cal F}\over \partial t_n\partial t_m}
=  - {\partial u_{n+1}\over\partial a_i}{\partial u_{m+1}\over\partial a_j}\
{\partial\over \partial \tau_{ij}}\log\ \vartheta_E( 0|\tau), \ \ \ \ \ \  n,m<N
}
with
\eqn\ucond{
u_n = {\p \CF\over \p t_{n-1}} =
{1\over n} {\rm res}_{P_+}\left(z^n {P_N'dz\over y}\right) =
{1\over n}\sum_{l=1}^N v_l^n = {1\over n}\langle\tr\ \Phi^n\rangle
}
justifying \defprep.

Our derivation of the renormalisation group equation almost repeats, or even
simplifies, the derivation from \gmmm. The main difference is that the
formalism of quasiclassical hierarchy is developed now at fixed hyperelliptic
co-ordinate $z$, while in \gmmm\ the role of such distinguished co-ordinate was
played by the co-ordinate on base torus $w$. In the last case the quasiclassical
flows were not deforming the geometry of the Seiberg-Witten curves (only the
generating differential had to be replaced), but the Losev-Shatashvili polynomials (the
analogs of \poluC) were never vanishing, being instead related to the generalised
Kontsevich model, or topological Landau-Ginzburg models \ref\lggkm{S.~Kharchev,
A.~Marshakov, A.~Mironov and A.~Morozov,
  {\it ``Landau-Ginzburg topological theories in the framework of GKM and
  equivalent hierarchies''}, hep-th/9208046,
  Mod.\ Phys.\ Lett.\  {\bf A8}, 1047 (1993)}.
Now we see, that algebraic $w$ has to be replaced by the transcendental $\Phi$.
The proper choice of local co-ordinate $z$ is suggested by the microscopic theory,
where it is already encoded in $\CF_{\rm UV}= \tr\ {\bf t} (\Phi)$,
see \fuv, and therefore it appears to be distinguished in the effective
functional \functnl. We can conclude therefore, that the quasiclassical
hierarchies arise as adequate language for the effective low-energy theories,
but to specify the details (the choice of proper basis, local co-ordinates etc)
one needs to turn directly to a microscopic theory.

\subsec{Instanton expansion with higher times}

Let us now test our results in the non-abelian case
against explicit calculations of the
first instanton corrections to the prepotential. We shall see immediately that
extracting even the first term in the instanton expansion of the Krichever
tau-function is a non-trivial task, and the direct evaluation using the
sum over partitions here is much more effective. The true application
of the quasiclassical tau-function is the analysis of the non-perturbative
effects such as the emergence of the  massless BPS particles at some points
of the moduli space of vacua, the surfaces of marginal stability and so on.
However, when we study
the extended Seiberg-Witten theory, the "new" directions
coming from the "times" $t_k$ which couple to the higher Casimirs,
are to be viewed as nilpotent (much like the generalised moduli space of
the topological string of B type on Calabi-Yau manifold \bmodel). We should not
attempt to give a physical meaning to the singularities which might occur at
finite values of $t_{k}$'s.
Note that adding matter to the ${\CN}=2$ theory effectively switches on the
times $t_{k}$. For example, if we add $N_f$ fundamental multiplets, we
get $t_{k} \sim {1\over {k-1}} \sum_{f} m_{f}^{-2k}$. In the theory with "real"
fundamental multiplets the prepotential has singularities when the charged matter becomes massless,
i.e. near $a \sim m_f$, which is near infinity for small $t$.

Let us start with the perturbative limit with the higher times switched on. It is
characterized by the degenerate differential \difff
\eqn\dfipert{
d{\Phi}_0 = {\bt}'''(z)dz - \sum_{l=1}^N{dz\over z-v_l}}
where the position of poles of degenerate holomorphic differentials
coincide with the perturbative values of the Seiberg-Witten periods
\eqn\av{
a_l = - {\rm res}_{v_l} ( zd\Phi_0 ) = v_l,\ \ \ \ l=1,\ldots,N
}
Integrating \dfipert,
\eqn\fipert{
\Phi_0 = {\bt}''(z) - \sum_{l=1}^N\log\left( z-a_{l}\right)
}
one gets the perturbative generating differential $dS_0=\Phi_0 dz$, satisfying
$$
{\partial dS_0 \over \partial a_l} =  {dz \over z-a_{l}}, \quad l = 1,\ \ldots , N
$$
$$
{\partial dS_0\over \partial t_k} =  kz^{k-1}dz,  \quad k>0
$$
and, in the rational case, it is easy to write the function
$$
S_0(x) = {\bt}^{\prime}(x) - \sum_{l=1}^N(x-a_l)\left(\log(x-a_l)-1\right)
$$
Equations
\eqn\SWpertS{
a^D_j = {\partial \CF_0\over\partial a_j} = S_0(a_j)
}
define the perturbative prepotential with switched on higher Casimirs to be
\eqn\prepert{
\CF_0 = \sum_{l=1}^N{\bt}(a_l) +
\ha \sum_{m\neq l} {\bF}(a_l-a_m)
}
to be the sum of the ultraviolet prepotential $\CF_{\rm UV}= \tr\ {\bf t} (\Phi)$
and the
perturbative $\CN=2$ prepotential, with the "Seiberg-Witten" function defined in \SWfun.

Consider now vanishing $B$-period \diffreq\ of the perturbative differential \dfipert\
$$
\int_{x_i^+}^{x_j^-} d\Phi_0 =0
$$
where $x_j^\pm = a_j \pm 2\sqrt{qS_j}+O(q^2)$ are positions of the branching points of
the curve \dcN\ in the vicinity of perturbative rational curve. Equation
$\Phi_0(x_i^+)=\Phi_0(x_j^-)$ in the first nonvanishing order in parameter of
instanton expansion $q$ gives
\eqn\Si{
S_{l} = {e^{2{\bt}^{\prime\prime}(a_{l})}\over\prod_{m\neq l}(a_{l}-a_{m})^2},\ \ \
l=1,\ldots,N
}
where the numeric coefficient is fixed, say, from comparison with
the Seiberg-Witten curve \swcurv\ on a small phase space.

Take now for simplicity the $U(2)$
gauge group. Then for the equation of the curve \dcN\
up to the second order in $q$ one has
\eqn\cuexp{\eqalign{ &
y^2 = \prod_{l=1,2}\left((z-a_l)^2-4qS_l + \ldots\right) = \cr
& =(z-a_1)^2(z-a_2)^2-4q\left(S_1(z-a_2)^2+S_2(z-a_1)^2\right) + O(q^2)
\cr }}
and
$$
{dz\over y} = {dz\over (z-a_1)(z-a_2)}\left(1+2q\left({S_1\over(z-a_1)^2}+
{S_2\over(z-a_2)^2}\right)\right) + O(q^2)
$$
One can also expand therefore \difff\ as $d\Phi=d\Phi_0+qd\Phi_1+O(q^2)$ with
\eqn\dfiq{\eqalign{ &
d\Phi_1=2\left({\bf t}'''(z) -{2z-a_1-a_2\over(z-a_1)(z-a_2)}\right)
\left({S_1\over(z-a_1)^2}+
{S_2\over(z-a_2)^2}\right)- \cr
& - {2(S_1-S_2)\over(a_1-a_2)(z-a_1)(z-a_2)}
\cr }}
normalized to ${\rm res}_{a_1}d\Phi_1={\rm res}_{a_2}d\Phi_1=0$
(we restrict ourselves now for ${\bf t}'''(z)=2t_2$ with only
nonvanishing $t_1, t_2\neq 0$).
Computation of residues of \dfiq\ at infinity gives
\eqn\resdfiq{\eqalign{ &
-\ha{\rm res}_\infty\left(z^2d\Phi_1\right) = a_1\delta a_1+a_2\delta a_2 +
S_1+S_2
\cr &
-{1\over 3}{\rm res}_\infty\left(z^3d\Phi_1\right) = a_1^2\delta a_1+a_2^2\delta a_2
+2\left(a_1S_1+a_2S_2\right)
\cr }}
Taking into account that \Ftres\ gives at linear order in $q$
\eqn\Fq{ -{1\over k+1}{\rm res}_\infty \left(z^{k+1}d\Phi_1 \right) =
\left.{\partial\CF\over\partial t_k}\right|_q = \sum_{l=1}^N a_l^{k-1}\delta a_l +
{\partial\CF_1\over\partial t_k}
}
where the first term in the r.h.s. arises as a contribution from the perturbative
prepotential \prepert\ due to instantonic renormalization of the relation between the
coefficients of the curve \dcN\ and the Seiberg-Witten periods
as $a_i \to a_i + q\delta a_i$ with
\eqn\deltaqa{\eqalign{ &
\delta a_1 = \left(2t_2-{1\over a_1-a_2}\right)S_1 =
{\rm res}_{a_1} \left( z d\Phi_1 \right)
\cr &
\delta a_2 = \left(2t_2+{1\over a_1-a_2}\right)S_2 =
{\rm res}_{a_2} \left( z d\Phi_1 \right)
\cr }}
Comparing \Fq\ with \resdfiq\ and taking into account \deltaqa,
one gets $\CF_1=\half\sum_l S_l$, what coincides with direct instantonic
calculus. Higher coefficients of the instantonic expansion of the quasiclassical
tau-function can be obtained in a similar way.

\newsec{Eguchi-Yang picture}

\ndt
In 1994, T.~Eguchi and S.~Yang have proposed \egu\
the following matrix model:
\eqn\matrm{Z_{N} (t) = \int {\CD}M \ {\exp} \ N {\tr} \left(
- 2 M \left( {\log} M - 1 \right) + \sum_{k>0} t_{k} M^{k} \right) }
for the description of the type $A$ ${\bC\bP}^{1}$ topological string, in the stationary
sector of Gromov-Witten theory (some generalisations for other target
spaces were studied in \tks). Their proposal is more general, it includes
the full phase space of the sigma model, i.e. the full set of gravitational
descendants of the cohomology of ${\bC\bP}^{1}$.

One can easily relate the matrix model \matrm\ to our variational problem \exeq.
We shall show that the proposal \egu\ is correct as far as the genus zero
part of the topological string is concerned, if one views \matrm\ as the integral
over the supermatrices of $(N_{+} | N_{-}) \times (N_{+} | N_{-})$ size.

Indeed, let us study the saddle point approximation to the matrix integral \matrm, in
the large $N$ limit.
As usual, we assume the eigen-values to form a distribution, with the continuous
density of the eigenvalues:
\eqn\dnse{{\rho}(x) = {1\over N}\langle {\tr}\ {\d} \left( x - M \right) \rangle}
The action of the matrix model \matrm\ together with the contribution of the measure
(the Vandermonde determinant) combine together nicely to the following effective
action functional of $\rho (x)$:
\eqn\effacmm{{\CV}_{\rm eff}[{\rho}]  = 2 \int_{x_{1} > x_{2}} \ dx_{1} dx_{2} {\rho}(x_{1}) {\rho}(x_{2})
{\rm log}( x_{1} - x_{2} )  + \int \ dx {\rho}(x) \left( - 2 x ({\rm log}(x) - 1) + \sum_{k>0} t_{k} x^{k} \right)
}
In the case of bosonic matrix model the density function ${\rho}
(x)$ is a non-negative function on the real line which has a compact support.
The obvious constraint for \dnse
\eqn\nrml{
\int \ dx {\rho}(x) = 1}
can be enforced with the help of a Lagrange multiplier:
$$
{\CV}_{\rm eff}[{\rho}] \to {\CV}_{\rm eff}[{\rho}] - a^{D}\left(\int \ dx
{\rho}(x) -1\right)
$$
The saddle point equation reads now:
\eqn\saddlmm{2\int\, d{\tilde x} {\rho}({\tilde x}) {\rm log}| x - {\tilde x}|
 - 2 x ({\rm log}(x) - 1) + \sum_{k>0} t_{k} x^{k}   = a^{D},
\qquad x \in {\rm supp}{\rho}}
Note that once we
shift $x \to x - a$, make an identification:
\eqn\idntc{f'(x) = {\rm sgn}(x) - {\rho}(x) }
and rescale the times $t_k$ and $a^{D}$ by an irrelevant here
factor of $2$\footnote{$^{\ddagger}$}{The factor of 2 in \matrm\ in front of the Eguchi-Yang
term is a consequence of the fact that matrix integral is rather related to the correlation
function bilinear in fermions, cf. e.g. with \comamo. In particular, this is related to the fact
that integration measure contains the {\it square} of the Vandermonde determinant.}, the equation \saddlmm\ becomes equivalent to \exeq.  It can be
shown by integrating \exeq\ by parts and choosing the real branch of the logarithm function.
However, the "density" ${\rho}(x)$ which corresponds to the saddle point of \dfunctnl\
is not non-negative. For example,  on the small phase space, for $a = 0$,
\eqn\fppp{
{\rho}(x) ={\rm sgn}(x)-{2\over{\pi}} {\rm arcsin} \left( {x  \over 2 e^{t_{1}}} \right)
}
which has a jump at $x=0$, where it goes from being positive to
being negative. Note also that for \fppp\ obviously:
\eqn\nrml{\int\ dx \ {\rho}(x) = 0}
(which is also consistent with \idntc\ and \fxda).

{}What could this negative density of eigenvalues possibly mean?
The hint comes from \nrml. Since the integral of the density of eigenvalues
is proportional to the trace of unit matrix, one can conclude that the
"matrix model" corresponding to the original model of random partitions \zuone\ must
have vanishing trace of the unit matrix. The natural venue for such matrices is the supermatrix model (see, e.g.
\ref\supema{ S.~A.~Yost,
 {\it ``Supermatrix models''}, hep-th/9111033,
  Int.\ J.\ Mod.\ Phys.\  {\bf A7} (1992) 6105}),
where the traces are replaced by supertraces. For the
$(N_{+}|N_{-}) \times (N_{+}|N_{-})$ matrices the supertrace of a unit operator is equal to $N_{+}-N_{-}$,
which can be both positive and negative.
\eqn\suprmt{M = \pmatrix{A & B \cr C & D}, \qquad {\rm str}M = {\tr}A - {\tr}D}
where $A$ and $D$ are $N_{+}\times N_{+}$ and $N_{-}\times N_{-}$ size hermitian
matrices with bosonic entries, respectively;
$B$ and $C$ are fermion-valued $N_{+}\times N_{-}$ and $N_{-} \times N_{+}$
dimensional complex matrices, $B = C^{\dagger}$.
The only subtlety in the definition of the measure ${\CD}M$ is the division by the
volume of the gauge group, which is done by the proper treatment of ghosts
(both bosonic and fermionic). The invariance of the measure and
the action (which is the same as \matrm\ with the replacement
$N{\tr} \to {1\over h} {\rm str}$, with the new "supermatrix Planck constant" $h$)
is such that both $A$ and $D$ can be diagonalised:
\eqn\diagad{ A \to {\rm diag}( x_{1}^{+}, \ldots , x_{N_{+}}^{+}) ,\ \ \
D \to {\rm diag}( x_{1}^{-}, \ldots , x_{N_{-}}^{-})}
We can view \diagad\ as a gauge-fixing condition. The corresponding Fadeev-Popov
determinant equals (note again its similarity with the (square of the)
free fermion correlator):
\eqn\fpv{\left( { \prod_{1\leq i < j\leq N_{+}} ( x^{+}_{i} - x_{j}^{+})
\prod_{1\leq i < j\leq N_{-}}
( x^{-}_{i} - x_{j}^{-}) \over  \prod_{1\leq i \leq N_{+}}\prod_{1\leq j\leq N_{-}}
( x_{i}^{+} - x_{j}^{-} ) }\right)^{2}}
 Now one can write for the density of eigenvalues \diagad:
 \eqn\corrden{{\rho}(x) = h \left( \sum_{i=1}^{N_{+}} {\d}( x - x^{+}_{i}) -
 \sum_{i=1}^{N_{-}}{\d}( x - x^{-}_{i})  \right) }
and consider it in the large $N$ limit, i.e.
when $h\to 0$ as $N_{\pm} \to \infty$, so that $h N_{\pm}$ are finite. The
logarithm of jacobian \fpv\ plus the Eguchi-Yang action are still given by
${1\over h^2}{\CV}_{\rm eff}[{\rho}]$, defined in \effacmm, but the supermatrix
density \corrden\ is automatically subjected to \nrml\ upon $N_{+} = N_{-}$, which
can be re-expressed in terms of times of the Toda hierarchy. For example,
on the small phase space:
\eqn\mn{hN_{\pm} \to  2{{\pi}- 2 \over {\pi}} e^{t_{1}} \ , }
while in general:
\eqn\mng{hN_{\pm} \to | x^{\pm}({\bf t}) - a |}
It would be nice to investigate our supermatrix model further, perhaps
extending the saddle point correspondence to the exact relation between the $\hbar$
and $1\over N_{\pm}$
(or $h$)-expansions. Perhaps it is worthwhile to study more general  supermatrix models, with
$N_{+} \neq N_{-}$.  It is also interesting to find out whether there is a relation of
our supermatrix model and the Dijkgraaf-Vafa matrix models, computing the superpotentials of
${\CN}=1$ supersymmetric gauge theories, as suggested in  \lmn\nikand.  In this paper
we have pointed out new features of the matrix models possibly relevant for the
${\CN}=2$ physics:
the non-positivity of eigenvalue densities, which could be related to the
block-matrix decompositions arising in the perturbative matrix model calculations
(cf. e.g.
with \ref\dgkv{R.~Dijkgraaf, S.~Gukov, V.~Kazakov and C.~Vafa,
  {\it ``Perturbative analysis of gauged matrix models''}, hep-th/0210238,
  Phys.\ Rev.\  {\bf D68} (2003) 045007},\ref\km{V.~Kazakov, A.~Marshakov,
  {\it ``Complex curve of the two matrix model and its tau-function''},
  hep-th/0211236,   J.\ Phys.\  {\bf A36} (2003) 3107} where such decomposition
  was performed for bosonic one- and two- matrix models).

\newsec{Conclusions and future directions}

In this paper we have solved the long-standing problem of extending
the prepotential of the effective ${\CN}=2$ theory to the "large phase space",
obtained by turning on the chiral perturbations
\eqn\opsuv{
{\Delta}{\CL} = \int\ d^4{\theta} \ {\bt} \left( {\bf\Phi} \right)
}
where ${\bt}$ is a gauge-invariant function. We have seen that for the
single-trace perturbations the resulting prepotential coincides with the
Krichever tau-function of the quasiclassical hierarchy. For the $U(1)$ theory
this is the dispersionless Toda chain hierarchy. For the $U(N)$ theory this
more general quasiclassical or universal Whitham hierarchy
(in the sense of \kriw) of the Toda-chain type (i.e. it corresponds to hyperelliptic
curve with two marked points $P_{\pm}$).

The derivation of the paper was based on  a particular identification of the
operators \opsuv\ with some cohomology classes of the moduli spaces ${\CM}_{k}$ of
gauge instantons, and their continuation on the partial compactifications
${\overline\CM}_{k}$. For the gauge theory with gauge group $U(N)$
the compactification ${\overline\CM}_{k}$ is the moduli space of
rank $N$ torsion free sheaves on $X = {\bC\bP}^{2}$ with the fixed trivialisation at the fixed
complex line $D = {\bC\bP}^{1}_{\infty}$ (this is also a moduli space of
noncommutative instantons \niksch). The operator \opsuv\ of the form:
\eqn\chrnl{\int\, d^4{\theta} \, {\tr}\, {\bf\Phi}^k }
is identified with
the $k$-th component of the Chern character ${\rm ch}({\CE})$
of the universal sheaf
$$
{\CE} \to {\overline\CM}_{k} \times X
$$
One can imagine other definitions of the ultraviolet observables \opsuv.
This discussion is parallel to the discussion of the $2$-observables in the
two dimensional sigma models \freckles. The integrable structure which we find with our
prescription is very suggestive.
Of course, the arguments \ref\donagi{R.~Donagi, E.~Witten, {\it "Supersymmetric Yang-Mills
theory and integrable systems."}, hep-th/9510101, \np{460}{1996}{299-334}} which show
that the moduli space of vacua
of ${\CN}=2$ supersymmetric gauge theory is a base of an algebraic
integrable system, based on electric-magnetic duality, do not apply
{\it per se} to the large phase space, spanned by the couplings $t_k$. However,
one can think, at least formally, of them as coming from the degeneration of the
standard moduli space of vacua of the theory with larger gauge group, more
matter and so on.

The appearance of quasiclassical hierarchies in the Seiberg-Witten theory
was anticipated long
time ago \gkmmm. The program of studying the
"large phase space" of the Seiberg-Witten theory was put forward in
\issues\ and some proposals about corresponding integrable hierarchies
were made in
\gkmmm\gmmm\ref\whith{
T.~Nakatsu, K.~Takasaki, {\it "Whitham-Toda hierarchy and ${\CN}=2$
supersymmetric Yang-Mills theory"}, hep-th/9509162, Mod. \pla{11}{1996}{157}\semi
M.~Mari{\~n}o, G.~Moore, {\it "The Donaldson-Witten function for
gauge groups of rank larger than one"}, hep-th/9802185, \cmp{199}{1998}{25-69}\semi
J.~Edelstein, M.~Mari{\~n}o, J.~Mas, {\it "Whitham hierarchies, instanton corrections
and soft supersymmetry breaking in N=2 SU(N) super Yang-Mills theory"}, hep-th/9805172,
Nucl.Phys. B541 (1999) 671}. The subtlety of
all these proposals is the contact term ambiguity.
In a sense it is a standard problem of reconstructing the
ultraviolet theory given the infrared one. We have shown in this paper,
that the quasiclassical hierarchy, in the basis of time-variables proposed
by microscopic instanton theory, solves the ambiguity problem exactly
in the way anticipated in \issues. For small $k,l$ the Losev-Shatashvili polynomials
can be also studied using the blowup techniques
\issues\ref\nakajimayoshioka{H.~Nakajima, K.~Yoshioka, {\it "Lectures on
instanton counting"}, math.AG/0311058}.

%\vskip 15pt
%\ndt
One should keep in mind the richness of physics which is related to the
random partition models of which we studied the thermodynamic limits: our results
can be easily extended to the five dimensional supersymmetric gauge theories
(which means going beyond the small phase space analysis of \mntt) where they
become the statements about the dispersionless hierarchies governing the melting
of the crystals with appropriate symmetries (thus generalizing the work of
\ref\andkolya{A.~Okounkov, N.~Reshetikhin,  math.CO/0107056 \semi
  A.~Okounkov, R.~Kenyon, S.~Sheffield, math-ph/0311005\semi
  A.~Okounkov, R.~Kenyon, math.AG/0311062});  the same fermion correlators and
  their dispersionless limits show up in the physics of one dimensional electrons
 \ref\pasha{E.~Bettelheim, A.~Abanov, P.~Wiegmann,  nlin.SI/0605006} ("physics
 of quantum wires"). Finally, one may study more seriously the analogues of  Douglas-Kazakov phase transition on the large phase space, and also incorporate  the multi-trace ultraviolet perturbations.  We hope to return to this discussion elsewhere.
\vskip 15pt
\noindent
{\bf Acknowledgements.}

We are grateful to Sergei Kharchev, Igor Krichever, Andrei Losev and  Andrei Okounkov for numerous discussions.

The work of AM was partially supported by Russian Ministry of Industry, Science and Technology under the contract 40.052.1.1.1112,
by the grants {\cyr RFFI} 05-02-17451, {\cyr NSh}--4401.2006.2 and
INTAS 05-1000008-7865, by the grant ANR-05-BLAN-0029-01
"Geometry and integrability in mathematical physics", and by the
NWO-{\cyr RFFI} program 047.017.2004.015
"Geometric Aspects of Quantum Theory and Integrable Systems".
The research of NN was partially supported by
European RTN under the contract 005104  "ForcesUniverse", by ANR
under the grants ANR-06-BLAN-3$\_$137168 "Structure of vacuum, topological
strings and  black holes", and ANR-05-BLAN-0029-01 "Geometry and integrability
in mathematical physics",  and by the grants {\cyr RFFI} 06-02-17382
and {\cyr NSh}--8065.2006.2. AM thanks Max-Planck-Institut f\"ur Mathematik, Bonn;
Institut des Hautes Etudes Scientifiques, Bures-sur-Yvette, and Laboratoire de
Math\'ematiques
et Physique Th\'eorique de l'Universit\'e de Tours, while NN thanks the Institute
for Advanced Study at Princeton, MSRI,
Department of Mathematics at UC Berkeley and
the New High Energy Theory Center at Rutgers University, for hospitality
during various stages of preparation of the manuscript.

\appendix{A}{Partitions and Chern characters}

\ndt
{\it Partitions and Young diagrams} (the standard reference
is \uchebnik).

\ndt
For each partition ${\l}$ \lp, its {\it size} $| {\l} | = {\l}_{1} + {\l}_{2} +
\ldots + {\l}_{{\ell}_{\l}}$ is a number of boxes in the so-called Young diagram
$Y({\l})$, which is a geometric way to represent partition. The length
${\ell}_{\l}$ is the number of columns.
Traditionally one draws Young diagrams as the collection of rows of boxes,
of the lengths ${\l}_{1}$, ${\l}_{2}$, etc. Here we draw an example of the
Young diagram, for the partition $\lambda = (5,2)$ with ${\ell}_\lambda=2$,
$|\lambda|=7$:
\vskip.5cm

\centerline{\vbox{\hbox{$\mbox{.27}{.27}\mbox{.27}{.27}\mbox{.27}{.27}
\mbox{.27}{.27}\mbox{.27}{.27}$}
\hbox{$\mbox{.27}{.27}\mbox{.27}{.27}$}}}
\ndt
\vskip.5cm
\ndt
and its profile function $f_\lambda(x)$ defined in \proff,
\Figx{14}{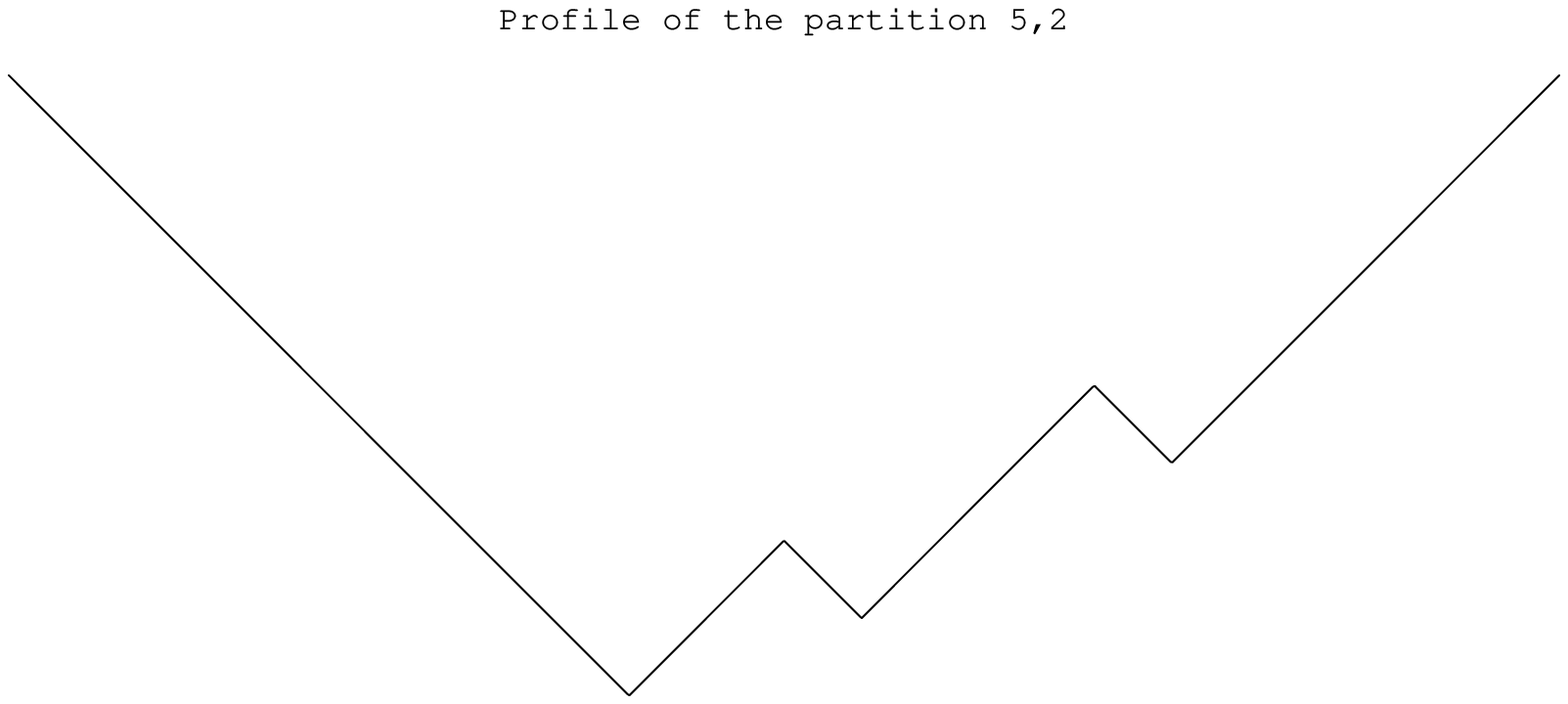}{}

\vskip.5cm
For partition $\l$ one introduces the {\it dual partition} ${\l}^{\prime}$, whose
Young diagram $Y({\l}^{\prime})$
is obtained by exchanging the rows and columns of $Y({\l})$.
For example, the partition,
\vskip.5cm

\hbox{\vbox{\hbox{}\hbox{dual to $(5,2)$,}\hbox{}\hbox{}\hbox{}} \

\vbox{\hbox{$\mbox{.27}{.27}\mbox{.27}{.27}\mbox{.27}{.27}\mbox{.27}{.27}\mbox{.27}{.27}$}
\hbox{$\mbox{.27}{.27}\mbox{.27}{.27}$}\hbox{}\hbox{}}\
\vbox{\hbox{}\hbox{}\hbox{\ , \ is the partition\ }\hbox{\qquad\ $(2,2,1,1,1)$:\ }\hbox{}\hbox{}}\
\vbox{\hbox{$\mbox{.27}{.27}\mbox{.27}{.27}$}
\hbox{$\mbox{.27}{.27}\mbox{.27}{.27}$}
\hbox{$\mbox{.27}{.27}$}
\hbox{$\mbox{.27}{.27}$}
\hbox{$\mbox{.27}{.27}$}}}

\vskip.5cm

\noindent
{}It is convenient to introduce a coordinate system, $(i,j)$, on  the Young diagram:
\eqn\sscrr{\eqalign{& 1 \leq j \leq {\l}_{i} \leq {\ell}_{{\l}^{\prime}}\ , \cr
& 1 \leq i \leq {\l}^{\prime}_{j} \leq {\ell}_{{\l}} \ . \cr}}
The coordinate $i$ labels rows, from top to bottom, while the coordinate $j$
labels columns, from left to right. For example, the partition $\lambda=(5,2)$ gets
coordinatised in the following way:
$$
\vbox{\hbox{\mathboxit{(1,1)}\mathboxit{(1,2)}\mathboxit{(1,3)}\mathboxit{(1,4)}\mathboxit{(1,5)}}
\hbox{\mathboxit{(2,1)}\mathboxit{(2,2)}}}
$$
For the box with the coordinates $(i,j)$, the {\it hook-length}
is defined as: $h_{(i,j)} = {\l}_{i} + {\l}^{\prime}_{j} - i - j  +1$.
In our example ${\l} = (5,2)$:
$$
\vbox{\hbox{\mathboxit{h_{(1,1)}=6}\mathboxit{h_{(1,2)} = 5}\mathboxit{h_{(1,3)}=3}\mathboxit{h_{(1,4)}=2}\mathboxit{h_{(1,5)}=1}}
\hbox{\mathboxit{h_{(2,1)}=2}\mathboxit{h_{(2,2)}=1}}}
$$
Finally, the Plancherel measure, which appears in \zuone, \planch\ is just the product
of the hook-length's
\eqn\hookpl{
{\bm}_{\l} = \prod_{{\sqx} \in {\l}}
{1\over{h_{\sqx}}}  = \prod_{i=1}^{{\ell}_{\l}} {( {\ell}_{\l} - i )! \over
( {\ell}_{\l} + {\l}_{i} - i )! }
\prod_{1 \leq i < j \leq {\ell}_{\l}} {{{\l}_{i} - {\l}_{j} + j - i}\over {j-i}}
= \prod_{i < j} {{{\l}_{i} - {\l}_{j} + j - i}\over {j-i}}
}
In our example:
$$
{\bm}_{(5,2)} = {1\over 1\cdot 1 \cdot 2 \cdot 2 \cdot 3\cdot 5 \cdot 6}
= {1! \cdot 0! \over 6! \cdot 2!} \cdot {4 \over 1}
= {4\over 1}{7\over 2}{8\over 3}{9\over 4}{10\over 5}{11\over 6}{12\over 7}\ldots
{3\over 1}{4\over 2}{5\over 3}\ldots
$$

\vskip15pt
\ndt
{\it Chern characters.}

\ndt
The {\it Chern character} of the partition ${\l}$ is a generating function
for the symmetric functions of the {\it eigen-values} ${\l}_{i} - i$:
\eqn\chrch{\eqalign{& {\rm ch}_{\l}( M , u ) = \
\sum_{i=1}^{\infty} e^{ u ( M + {\half} - i +{\l}_{i})}\ = \  {e^{M u}
\over e^{u\over 2} - e^{-{u\over 2}}} +
\sum_{i=1}^{\infty} e^{u ( M + {\half} - i )} ( e^{u {\l}_{i}} -1) \  = \cr
& \qquad\qquad\  =\ {1\over u} \ +
M + u \left( {M^{2} \over 2} - {1\over 24} + \vert {\l} \vert \right)
+ \cr
& \qquad\qquad\qquad\qquad\qquad\ + {u^{2} \over 2} \left( {M^{3} \over 3
} - {M \over 12} + 2 M \vert {\l} \vert + \sum_{i} {\l}_{i}
\left( {\l}_{i} - 2i +1 \right) \right) + \ldots  \cr
}}
the second equality being valid when ${\Re}{\rm e} u > 0$.
In the gauge theory the main object is the Chern character of the
``universal sheaf'' ${\CE}$,
which is related to \chrch\ in the following simple way:
\eqn\gauch{\eqalign{& {\rm ch}({\CE}) = \left( e^{{\hbar}u\over 2} -
e^{-{\hbar u \over 2}} \right)
{\rm ch}_{\l} \left( {a\over {\hbar}} , {\hbar}u  \right)=  \cr
& \qquad\qquad = e^{u a} \left( 1 + ( 1 - e^{-u {\hbar}}) \sum_{i=1}^{\infty}
e^{u {\hbar}( 1 - i   ) } ( e^{u {\hbar}{\l}_{i}} - 1 ) \right)=\cr
& \qquad\qquad\qquad\qquad = \ \sum_{k=0}^{\infty}\
{u^k\over k!}\ {\rm ch}_{k}( a, {\l}) \cr}}
It is the components $ {\rm ch}_{k}( a, {\l})$ of Chern character \gauch\ that
enter the formula for the statistical weight of the partition $\l$ in the
ensemble \zuone.
For example:
\eqn\lowch{\eqalign{& {\rm ch}_{0}(a, {\l})  = 1 \cr
& {\rm ch}_{1}( a, {\l} ) = a \cr
& {\rm ch}_{2} ( a, {\l}) = a^2  +
2{\hbar}^{2} \vert {\l} \vert \cr
& {\rm ch}_{3}( a, {\l}) = a^{3} + 6{\hbar}^{2} a \vert {\l} \vert +
3{\hbar}^{3} \sum_{i}
{\l}_{i} ( {\l}_{i} + 1 - 2i ) \cr}}
One can also write for the Chern characters
\eqn\chernfo{\eqalign{ &
{\rm ch}_{k}( a, {\l}) = a^k  +\sum_{i=1}^{\infty}
\left( (a + {\hbar} ( {\l}_{i} - i +1 ))^k - (a +{\hbar}
( {\l}_{i} - i  ))^k\right.- \cr
& \qquad\qquad\qquad\qquad -\left.
(a + {\hbar} (1-i))^k+(a - {\hbar}i)^k\right) = \cr
& = \sum_{i=1}^{\infty}
\left( (a + {\hbar} ( {\l}_{i} - i +1 ))^k - (a +{\hbar}
( {\l}_{i} - i  ))^k\right)
\cr}}
which is useful to make link with the fermionic formalism.

\footatend\vfill\supereject\immediate\closeout\rfile\writestoppt
\baselineskip=14pt\centerline{{\bf References}}\bigskip{\frenchspacing%
\parindent=20pt\escapechar=` \input refs.tmp\vfill\eject}\nonfrenchspacing

\bye